\documentclass[prd,showpacs,floatfix,nofootinbib]{revtex4}
\usepackage{bm}
\usepackage{amssymb}
\usepackage{graphicx}
\usepackage{epstopdf}

\begin{document}
\title{Computational Efficiency of Frequency-- and Time--Domain Calculations of Extreme Mass--Ratio Binaries: Equatorial Orbits}
\author{Jonathan L.~Barton$^1$, David J.~Lazar$^1$, Daniel J.~Kennefick$^2$, Gaurav Khanna$^3$, and Lior M. Burko$^{1,4}$ }
\affiliation{$^1$ Department of Physics, University of Alabama in Huntsville, Huntsville, Alabama 35899, USA \\
$^2$ Department of Physics, University of Arkansas, Fayetteville, Arkansas 72701, USA\\
$^3$ University of Massachusetts at Dartmouth, North Dartmouth, Massachusetts 02747, USA\\
$^4$ Center for Space Plasma and Aeronomic Research, University of Alabama in Huntsville, Huntsville, Alabama 35899, USA}
\date{Draft of April 7, 2008}
\begin{abstract}
Gravitational waveforms and fluxes from extreme mass--ratio inspirals can be computed using time--domain methods with accuracy that is fast approaching that of frequency--domain methods. We study in detail the computational efficiency of these methods for equatorial orbits of fast spinning Kerr black holes, and find the number of modes needed in either method ---as functions of the orbital parameters--- in order to achieve a desired accuracy level. We then estimate the total computation time and argue that for high eccentricity orbits the time--domain approach is more efficient computationally. We suggest that in practice low--$m$ modes are computed using the frequency--domain approach, and high--$m$ modes are computed using the time--domain approach, where $m$ is the azimuthal mode number.
\end{abstract}
\pacs{04.25.Nx, 04.30.Db, 04.30.-w}
\maketitle

\section{Introduction and summary}

Compact stellar mass objects emit gravitational waves in the good sensitivity band of the planned Laser Interferometer Space Antenna (LISA) during their last year of inspiral into a supermassive black hole. The gravitational waveforms of EMRIs, extreme mass ratio insiprals, are therefore of interest, and when observed may shed light on the spacetime of the central black hole, including testing the Kerr hypothesis \cite{gair08}, in addition to being a sensitive tool to measure the central black hole's parameters. This possibility makes such sources of gravitational waves extremely interesting, despite the relatively low number of sources expected during the lifetime of the LISA mission. Specifically, tens to hundreds of such sources are expected to redshifts of $z\sim 0.5$--$1$ \cite{gair04}. In addition, the solution of the two body problem in general relativity remains challenging, and EMRIs orbits and waveforms correspond to its solution in the extreme mass ratio limit. 

Construction of theoretical templates is important both for detection of EMRIs gravitational waves and for accurate parameter estimation. Numerical waveforms can be constructed using the frequency--domain (FD) or the time--domain (TD) approaches. The former approach has been developed to very high accuracy, and is considered robust and accurate. On the other hand, advances to the TD approach have been hindered first by the success of the FD approach \cite{glampedakis02,hughes00}, and by the crudity of the initial attempts to evolve numerically the fields coupled to a point-like source with the Teukolsky equation \cite{lopez_aleman,khanna04}. The breakthrough in the accuracy of TD solutions of the inhomogeneous Teukolsky equation, i.e., the 2+1D solution of the Teukolsky equation coupled to a point mass, was recently achieved in \cite{burko-_khanna_07,pranesh}. For the first time, it was shown that TD calculations can be as accurate as FD calculations. The TD method of \cite{burko-_khanna_07} was improved with the introduction of the ``discrete delta" model of the source \cite{pranesh} and an appropriate low pass filter that makes the discrete delta useful also for eccentric or inclined orbits \cite{pranesh2}. Specifically, correlation integrals of gravitational waveforms done for the same system in the FD and TD approaches  show that the two agree to a high level \cite{pranesh2}. One may therefore argue that the two methods are comparable in the results they are capable of producing. We therefore contend that the viewpoint that the TD solution of the inhomogeneous Teukolsky equations is far from being competitive  from the FD solution can no longer be supported. However, we believe that one should not seek competition of the two approaches, but rather how they complement each other, as either method has non-overlapping strengths. In order to achieve this goal one needs to compare the computational efficiency of the two approaches. 

The question of the efficiency and the computation time with which the results are obtained remains an open question though. 
The common wisdom is that the FD computation is more effective computationally than  its TD counterpart: 
FD approaches are particularly convenient when the system ---and the emitted gravitational waves--- exhibits a discrete set of frequencies. Indeed, as shown by Schmidt~\cite{schmidt} and by Drasco and Hughes~\cite{drasco-hughes}, all bound Kerr orbits have a simple, discrete spectrum of orbital frequencies. However, generic orbits and in particular high-eccentricity orbits, although in principle amenable to a Fourier decomposition and a FD construction of the waveforms, require the summation of many terms in the Fourier series. This problem limits the accuracy and increases the computation time in FD calculations. While this statement, or similar ones, appear in the literature \cite{glampedakis02}, it has not been quantified. 

The motivation of this paper is to study the relative computational efficiency of FD and TD codes for the solution of the inhomogeneous Teukolsky equation. Specifically, we study the question of how much computational time is needed to find the fluxes of energy and angular momentum to infinity from an eccentric and equatorial orbit around a fast spinning Kerr black hole. We restrict the analysis here to equatorial orbits. Analysis of equatorial orbits is non-trivial, and teaches much of the method and properties also of non-equatorial orbits. Analysis of non equatorial orbits increases considerably the volume of the parameter space, and we leave its study to the future. We do, nevertheless, comment on the analysis of generic orbits.  

The estimation of the total needed computation time for TD codes is relatively simple: after the desired accuracy level is set, one needs to estimate the number of azimuthal $m$ modes (associated with the $\phi$ orbital angular frequency) required for the total sum over $m$ modes to achieve the desired accuracy, and then compute each $m$ mode at the same accuracy level. The number of needed $m$ modes is not hard to estimate, because the partial fluxes approach a geometric progression in $m$ for large $m$ values, with the asymptotic factor between two successive partial fluxes depending on the eccentricity of the orbit. After the needed number of $m$ modes is determined, one can use the TD code of \cite{burko-_khanna_07} (possibly with the improvements included in \cite{pranesh,pranesh2}) to calculate the partial fluxes and sum over them. Using the asymptotic geometric progression structure one may also estimate the error associated with the summation over the partial fluxes, and verify that the desired accuracy level is indeed achieved. This is done in Section \ref{m_modes}. The TD calculation of the individual partial fluxes can be done more efficiently than a straightforward calculation of the fields to very late times and great distances, that are required for an accurate estimate of the fluxes at infinity. Specifically, one can make use of the peeling properties of the Weyl scalars at great distances, and fit the fields at finite distances to that behavior. One may then use the fitted behavior to extract the behavior at infinity. This is done in Appendix \ref{appendixA}. We find this method to be cleaner and better motivated physically than the fit done in \cite{pranesh}, that uses a general two--parameter fit function of a form which is not strongly motivated physically. 

The estimation of the total needed computation time for FD codes starts similarly to the TD analysis: specifically, one needs to first determine the number of $m$ modes needed for the determined accuracy level. This can be done in a similar way to how it is done in the TD. The calculation of each $m$ mode in the FD approach requires the summation over a number of $k,n,\ell$ modes, that correspond to the radial and  angular (about the equatorial plane) orbital angular frequencies and the radiative multipole $\ell$. Because we focus attention on equatorial plane orbits, the $n,\ell$ modes trivialize, and in practice only the $k$ modes are of importance. We find in Section \ref{k_mode} that the number of $k$ modes that one needs to sum over to achieve the desired accuracy level increases with the eccentricity and with the corresponding value of $m$ in an intricate manner. 

Next, in Section \ref{comp_time} we estimate the computation time of each $k$ mode, and find it is a function of $k$. The behavior of the computation time of the $k$ mode is found to have a rather intricate dependence of the value of $k$. We use this behavior to approximate the computation time over a sum of $k$ modes up to some $k_{\rm max}$, and argue it is approximately quadratic in $k_{\rm max}$ for high $k_{\rm max}$ values. Then, in Section \ref{compare} we find the total computation time for all the FD modes on the same machine on which we perform the TD computation, and compare the two. We find that the FD code is more efficient at low $m$ values (for a twofold reason: because those require few $k$ modes, and for low $m$ values each $k$ mode takes less time to compute), but the computation time increases rapidly with the value of $m$. We find the growth rate of the computation time with the mode number $m$ to increase as a power of $m$, with the value of the exponent increasing with the eccentricity. Finally, we estimate that for generic orbits the two methods become comparable already for moderately high values of the eccentricity, in the range of $\epsilon\sim 0.6$--$0.7$. Higher eccentricity orbits are more efficiently calculated using the TD approach. Even when the calculation of the sum of all $m$ modes is done more efficiently using one method over the other, one may still compute the low $m$ modes using the FD approach and the high $m$ modes using the TD approach. Such a hybrid method may prove to be the most efficient computationally. One should also consider the fact that the total needed number of $k$ modes is an empirically found number. When the full parameter space is mapped, one may tabulate the numbers of required modes as functions of the orbital parameters. Until this is done, extra computation time is needed to find the number of needed modes, including the  computation of some modes that are found {\em a posteriori} to be unneeded. No similar problem occurs in the TD approach.

\section{Summing over the $m$ modes}\label{m_modes}
To obtain the sum over all $m$ modes we need to obtain results for high $m$--numbers, and find a way to determine (i) how many $m$ modes we need to sum over to get the total flux to a certain pre-determined accuracy, and (ii) estimate the error in neglecting all the higher $m$ modes. In the case of circular and equatorial orbits, Finn \& Thorne \cite{finn_thorne_00} show that  (see \cite {finn_thorne_00} for more details and for definitions) 
$${\dot E}_m=\frac{2(m+1)(m+2)(2m+1)!\, m^{2m+1}}{(m-1)[2^m\, m!\, (2m+1)!!]^2}\,\eta^2{\tilde \Omega}^{2+2m/3}\,{\dot{\cal E}}_{\infty\, m}$$
which has the nice property that 
${\dot E}_{m+1}/{\dot E}_m\longrightarrow {\rm const}$ as $m\to\infty$. (Notably, this property depends on the factor of ${\tilde\Omega}^{2+2m/3}$.) We may therefore approximate the sum over infinitely many modes by taking the series to be a geometric progression. 

\begin{figure}
\input epsf
\includegraphics[width=11.0cm]{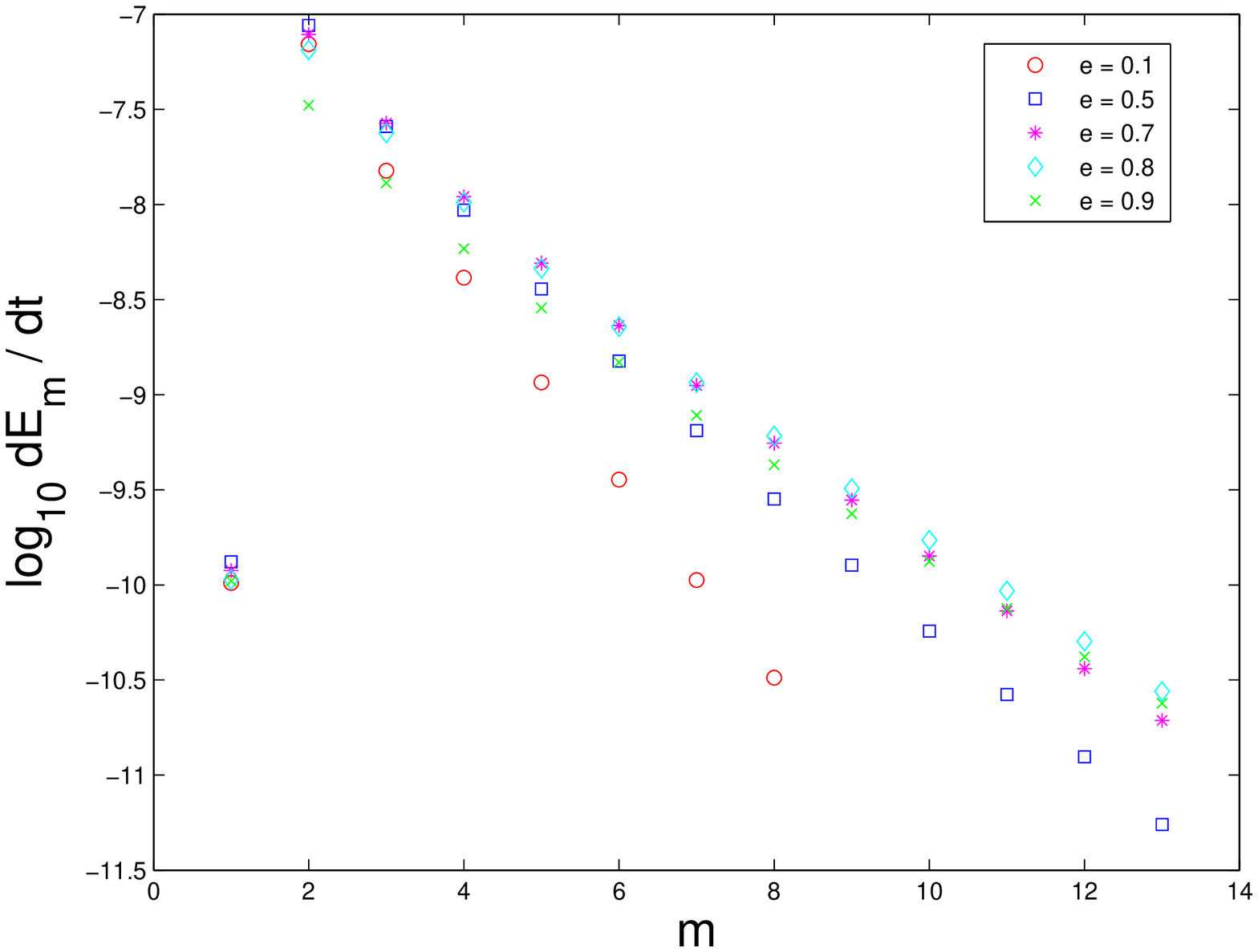}
\caption{The flux in the $m$ mode as a function of $m$ for various values of the eccentricity $e$. The black hole's spin is 
$a/M=0.9$ and the orbit's semilatus rectum is $p/M=4.64$.}
\label{fig1}
%
\includegraphics[width=11.0cm]{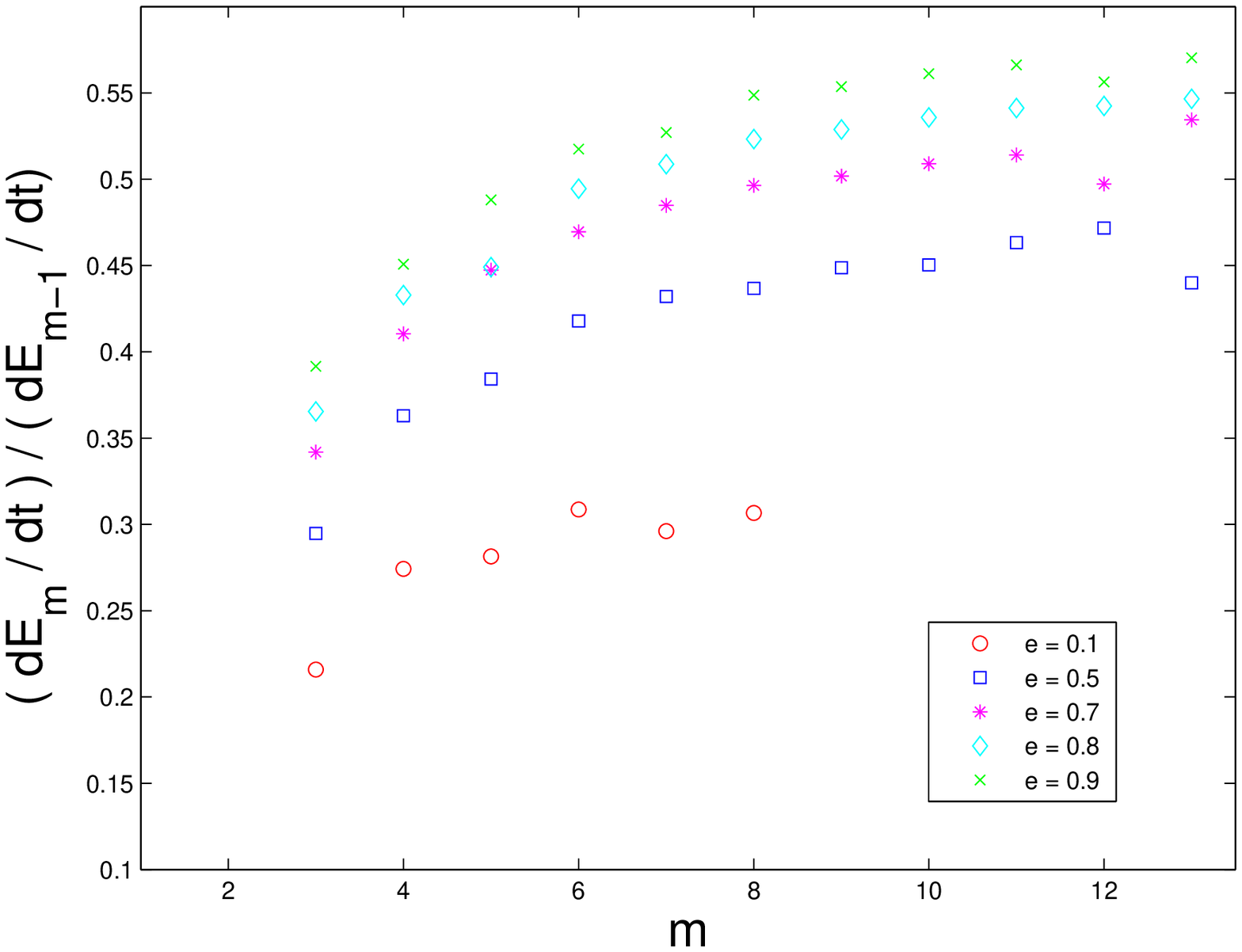}
\caption{The ratio of the relative contribution of the $m$ mode to the partial-sum (up to $m$) flux compared with the preceding mode, as a function of $m$, for various values of the eccentricity $e$. Same parameters as in Fig.~\ref{fig1}.}
\label{fig2}
\end{figure}

For eccentric orbits we no longer have the Finn--Thorne formula, but we can do numerical experiments to test whether the same results holds also for such orbits. We first show in Table \ref{table5} the average fluxes obtain for a number of eccentric orbits (with a fixed value of the semilatus rectum $p$) as a function of the mode $m$. The data presented is Table \ref{table5} is intentionally coarse, and is not intended to be more accurate than at the 5--10\% level. Below, we also present similar data with higher accuracy, where the latter are needed. We plot in Fig.~\ref{fig1} ${\dot E}_m$ as a function of $m$ for several eccentric orbits. We see that for all values of $\epsilon$, the drop off rate is exponential in $m$ for large values of $m$. Notably, the highest flux is in the fundamental mode, $m=2$, and is highest for $\epsilon\sim 0.5$. This result is indeed expected: The radiated power in gravitational waves averaged over one period for a point mass $\mu$ in a Keplerian orbit around a Schwarzschild black hole of mass $M$ is given by the quadrupole formula to be \cite{peters_mathews_63}
$$
\langle P\rangle=\frac{32}{5}\frac{\mu^2M^2(M+\mu)}{a^5(1-\epsilon^2)^{7/2}}\left(1+\frac{73}{24}\epsilon^2+\frac{37}{96}\epsilon^4\right)\, .
$$
Recalling that the semimajor axis $a$ is related to the semilatus rectum $p$ by $p=a(1-\epsilon^2)$, we find that for fixed $p$, 
$$\langle P\rangle \sim (1-\epsilon^2)^{3/2}\left(1+\frac{73}{24}\epsilon^2+\frac{37}{96}\epsilon^4\right)\, ,
$$
which has a maximum at $\epsilon= 0.465$ \cite{finn}. As the partial flux in any other mode is suppressed by over an order of magnitude compared with the fundamental mode ($m=2$), this result for the total flux is carried over to the fundamental mode. Notably, the larger $m$, we find numerically that the larger the eccentricity for which most flux is obtained.  

\begin{table}[h]
 \caption{Average (over one period) fluxes per unit mass extracted at $r=100M$ for a central black hole with $a/M=0.9$, and a prograde eccentric orbit of semilatus rectum $p=4.64M$, for various values of the eccentricity $\epsilon$. For each mode $m$, we show in bold print the mode that corresponds to the eccentricity for which the highest flux is achieved. Notice that the data presented in this Table are rather coarse, and are only intended to demonstrate the overall behavior. The accuracy level of data in this Table is at the 5--10\% level. Fine details are studied below. }
  \centering
     \begin{tabular}{|c||c|c|c|c|c|} \hline
   $|m|$ & $\epsilon=0.1$ & 0.5 & 0.7 & 0.8 & 0.9 
   \cr \hline \hline
   1 & $1.03\times 10^{-6}$ & ${\bf 1.32\times 10^{-6}}$ & $1.19\times 10^{-6}$ & $1.07\times 10^{-6}$ &    $1.05\times 10^{-6}$ \cr \hline
     2 & $6.96\times 10^{-4}$ & ${\bf 8.73\times 10^{-4}}$ & $7.84\times 10^{-4}$ & $6.49\times 10^{-4}$ & $3.32\times 10^{-4}$  \cr \hline
        3 & $1.50\times 10^{-4}$ & $2.57\times 10^{-4}$ & ${\bf 2.68\times 10^{-4}}$ & $2.37\times 10^{-4}$ &  $1.30\times 10^{-4}$\cr \hline
           4 & $4.12\times 10^{-5}$ & $9.34\times 10^{-5}$ & ${\bf 1.10\times 10^{-4}}$ & $1.03\times 10^{-4}$ & $5.86\times 10^{-5}$ \cr \hline
              5 & $1.16\times 10^{-5}$ & $3.59\times 10^{-5}$ & ${\bf 4.92\times 10^{-5}}$ & $4.61\times 10^{-5}$ &   $2.86\times 10^{-5}$ \cr \hline
                 6 & $3.58\times 10^{-6}$ & $1.50\times 10^{-5}$ & ${\bf 2.31\times 10^{-5}}$ & $2.28\times 10^{-5}$ & $1.48\times 10^{-5}$  \cr \hline
                    7 & $1.06\times 10^{-6}$ & $6.48\times 10^{-6}$ & $1.12\times 10^{-5}$ & ${\bf 1.16\times 10^{-5}}$ & $7.80\times 10^{-6}$  \cr \hline
                       8 & $3.25\times 10^{-7}$ & $2.83\times 10^{-6}$ & $5.56\times 10^{-6}$ & ${\bf 6.07\times 10^{-6}}$ & $4.28\times 10^{-6}$  \cr \hline
                    9 &  & $1.27\times 10^{-6}$ & $2.79\times 10^{-6}$ & ${\bf 3.21\times 10^{-6}}$ & $2.37\times 10^{-6}$  \cr \hline
                 10 &  & $5.72\times 10^{-7}$ & $1.42\times 10^{-6}$ & ${\bf 1.72\times 10^{-6}}$ & $1.33\times 10^{-6}$  \cr \hline
              11 &  & $2.65\times 10^{-7}$ & $7.30\times 10^{-7}$ & ${\bf 9.31\times 10^{-7}}$ & $7.53\times 10^{-7}$  \cr \hline
            12 &  & $1.25\times 10^{-7}$ & $3.63\times 10^{-7}$ & ${\bf 5.05\times 10^{-7}}$ & $4.19\times 10^{-7}$   \cr \hline
          13 &  & $5.50\times 10^{-8}$ & $1.94\times 10^{-7}$ & ${\bf 2.76\times 10^{-7}}$ &  $2.39\times 10^{-7}$ \cr \hline  \hline           
   \end{tabular}
\label{table5}
\end{table}

We next consider the ratio of the relative contribution of the $m$ mode to the partial-sum (up to $m$) flux compared with the preceding mode, as a function of $m$. An asymptotic drop off of the flux corresponds to this ratio approaching a constant value as $m\to\infty$. In Fig.~\ref{fig2} we plot this ratio as a function of $m$ for various values of the eccentricity.

\begin{figure}
\input epsf
\includegraphics[width=11.0cm]{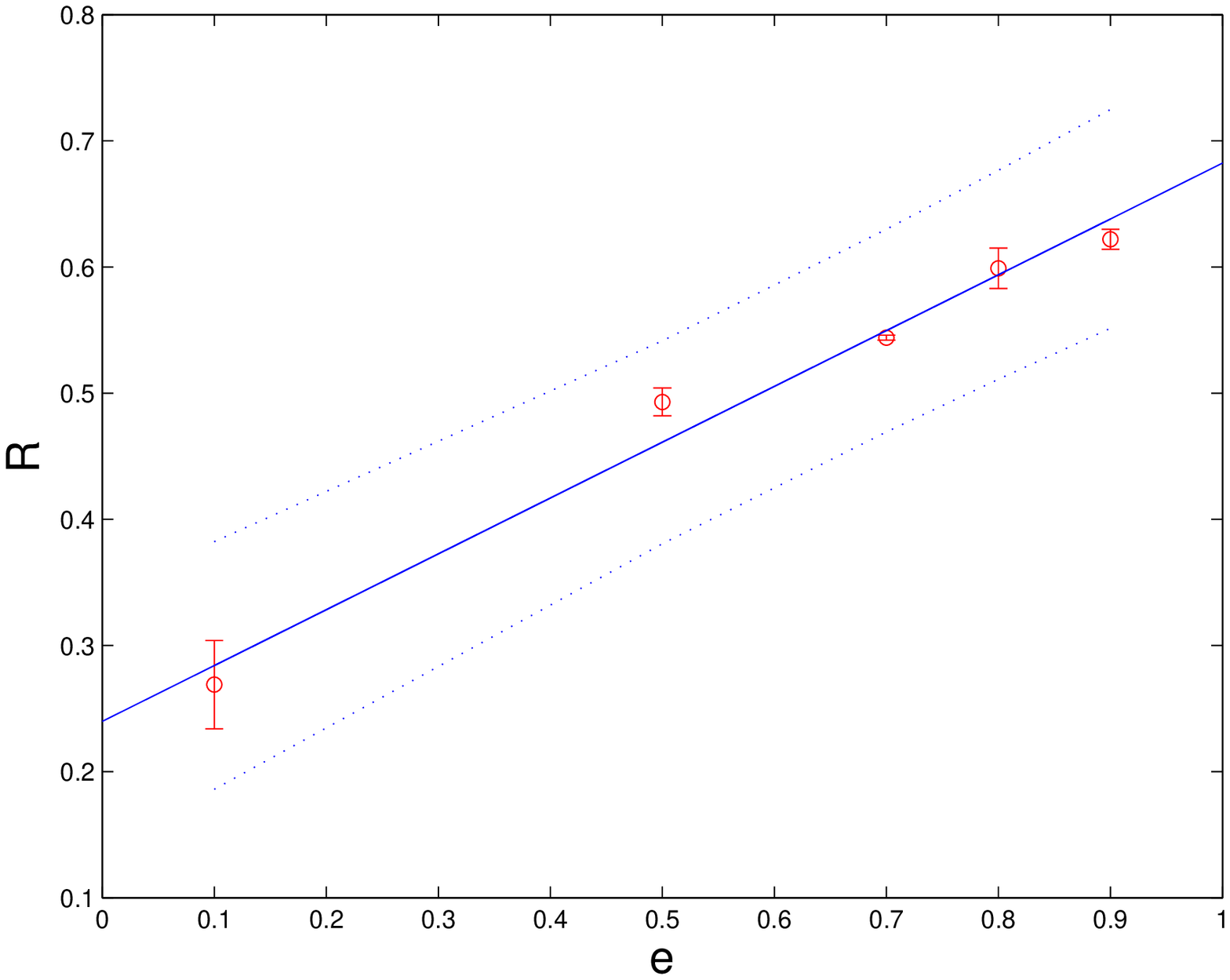}
\caption{The asymptotic ratio $R$ as a function of the eccentricity $e$. The individual error bars are shown, in addition to the best fit curve (solid) and $3\sigma$ confidence curves (dotted).}
\label{fig3}
\end{figure}

Lastly, we extrapolate the curves in Fig.~\ref{fig2} to find the asymptotic value for the ratio, and plot it in Fig.~\ref{fig3} as a function of the eccentricity $\epsilon$. The best fit curve is given by $R=0.24+0.44\,\epsilon$ 
with a correlation coefficient $R^2=0.980438$. 
 
\begin{figure}
\input epsf
\includegraphics[width=11.0cm]{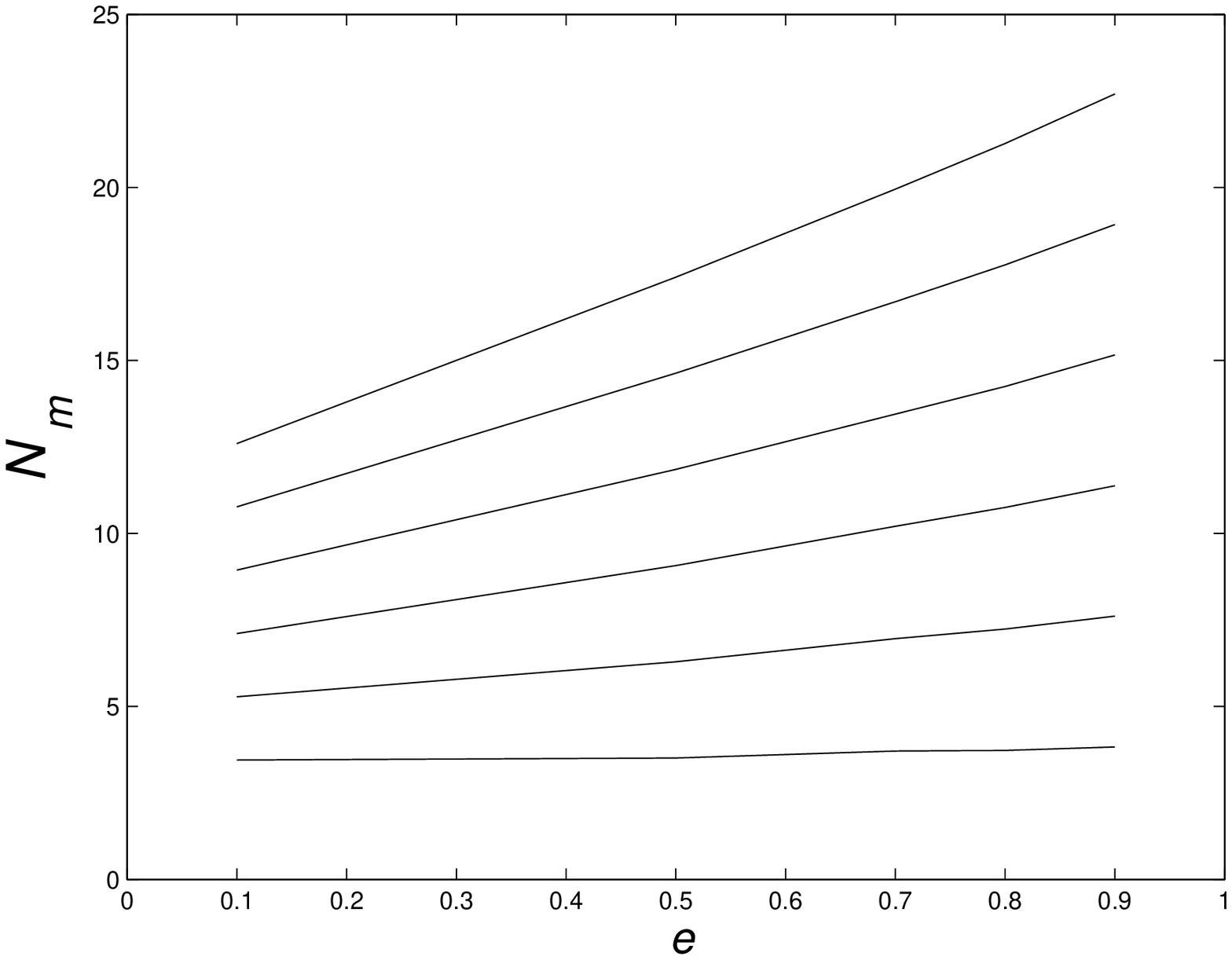}
\caption{The number of $m$-modes necessary for the sum over $m$-modes to have a certain accuracy. We plot contour curves at a fixed level of accuracy, showing the number of $m$-modes needed as a function of the eccentricity $e$. The contours are at accuracy levels of $10^{-1},10^{-2},10^{-3},10^{-4},10^{-5}$, and $10^{-6}$.}
\label{fig4}
\end{figure}

As ${\dot E}_m$ behaves asymptotically like a geometric progression, we can approximately sum over all modes (provided sufficiently many modes are calculated, so that the sequence of partial fluxes already converges approximately to the asymptotic behavior). Specifically, calculating the sequence
${\dot E}_1,{\dot E}_2,{\dot E}_3,\cdots ,{\dot E}_{n-1},{\dot E}_n$, we can calculate the partial sums 
$S_n:=\sum_{m=1}^n{\dot E}_m$, and the remainder is approximated by $R_n\sim {\dot E}_nR/(1-R)$, where $R$ is the asymptotic ratio evaluated above.  One can then approximate the total flux by 
${\dot E}_{\rm Total}\sim S_n+R_n$, and use $\Delta_n = R_n/{\dot E}_{\rm Total}$ as a measure for the 
error. We can therefore answer questions such as ``how many modes $m$ do we need to sum over to obtain a desired accuracy level?" (assuming each mode individually has the desired accuracy level). We show this in Fig.~\ref{fig4}, that shows contour curves at a fixed accuracy level on the number of $m$-modes--eccentricity plane.

\begin{figure}
\input epsf
\includegraphics[width=11.0cm]{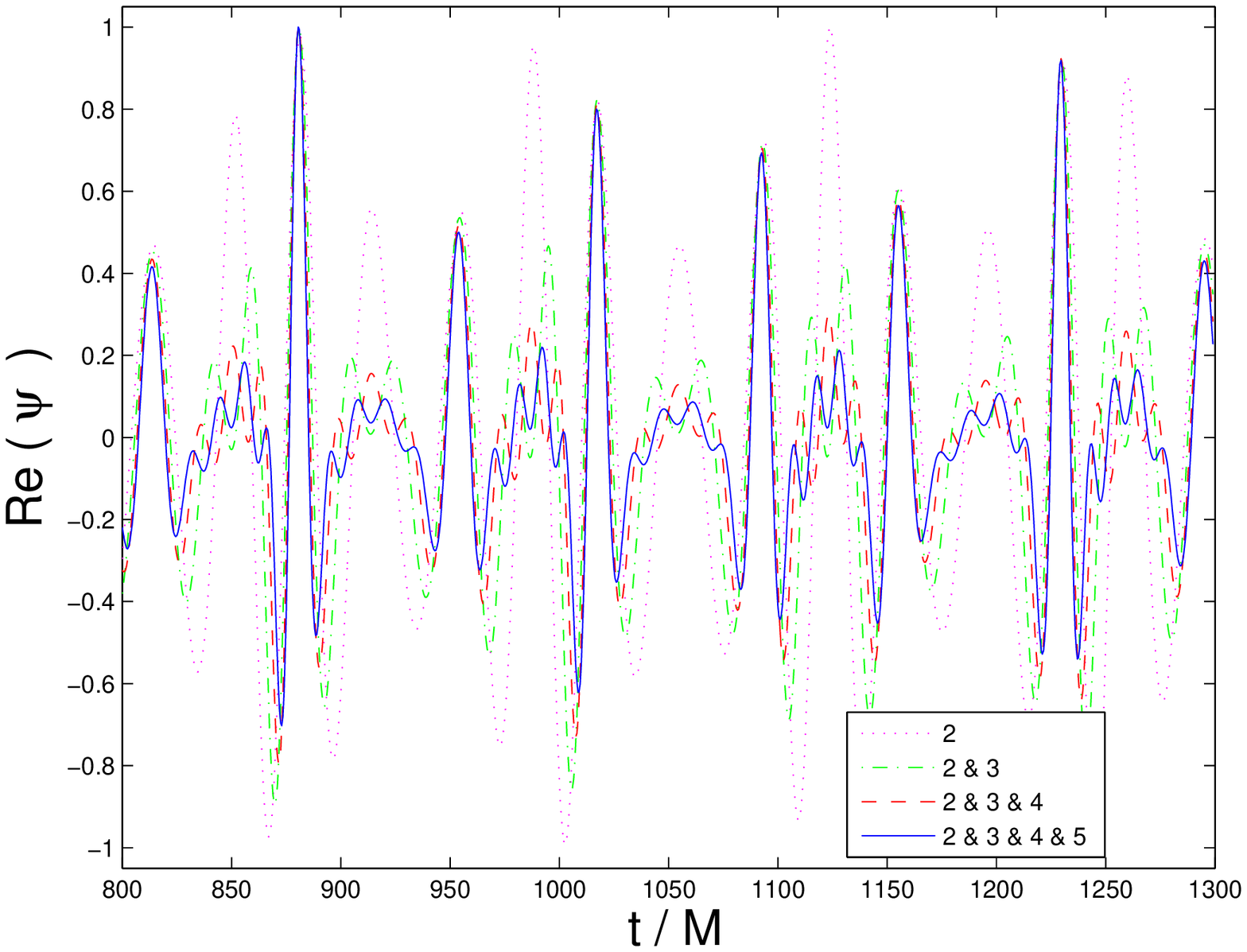}
\caption{The real part of the waveform on the equatorial plane at $r/M=625$ for $a/M=0.9$, $p/M=4.64$, and $\epsilon=0.1$ for $\psi_{m=2}$ (dotted), $\psi_2+\psi_3$ (dash-dotted), $\psi_2+\psi_3+\psi_4$ (dashed) and  $\psi_2+\psi_3+\psi_4+\psi_5$ (solid), as a function of $t/M$. }
\label{fig5re}
\includegraphics[width=11.0cm]{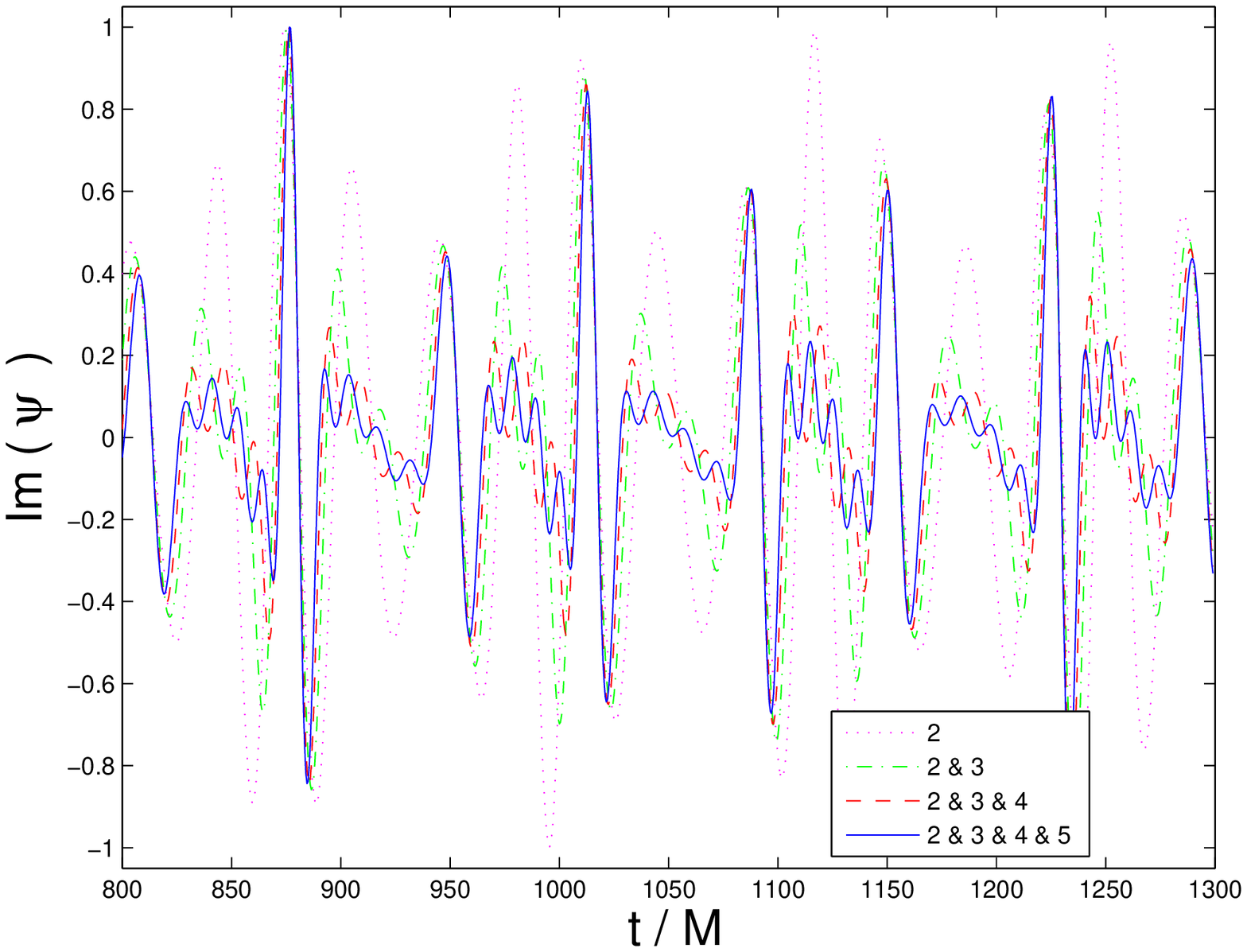}
\caption{The imaginary part of the waveform on the equatorial plane at $r/M=625$ for $a/M=0.9$, $p/M=4.64$, and $\epsilon=0.1$ for $\psi_{m=2}$ (dotted), $\psi_2+\psi_3$ (dash-dotted), $\psi_2+\psi_3+\psi_4$ (dashed) and  $\psi_2+\psi_3+\psi_4+\psi_5$ (solid), as a function of $t/M$. }
\label{fig5im}
\end{figure}

Based on Fig.~\ref{fig4}, to determine the energy flux to 10\% accuracy requires 4 modes at all values of the eccentricity $\epsilon$. We may therefore find the approximate total waveform by summing over the 4 modes of greatest energy flux. For $a/M=0.9$, $p/M=4.64$, and $\epsilon=0.1$, we show in Figs.~\ref{fig5re} and \ref{fig5im} the waveform obtained when summing over these 4 modes, specifically $m=2,3,4,5$. Indeed, the phase information is captured well by the sum of these 4 modes, and is no longer changing significantly by addition of more modes.

\begin{figure}
\input epsf
\includegraphics[width=11.0cm]{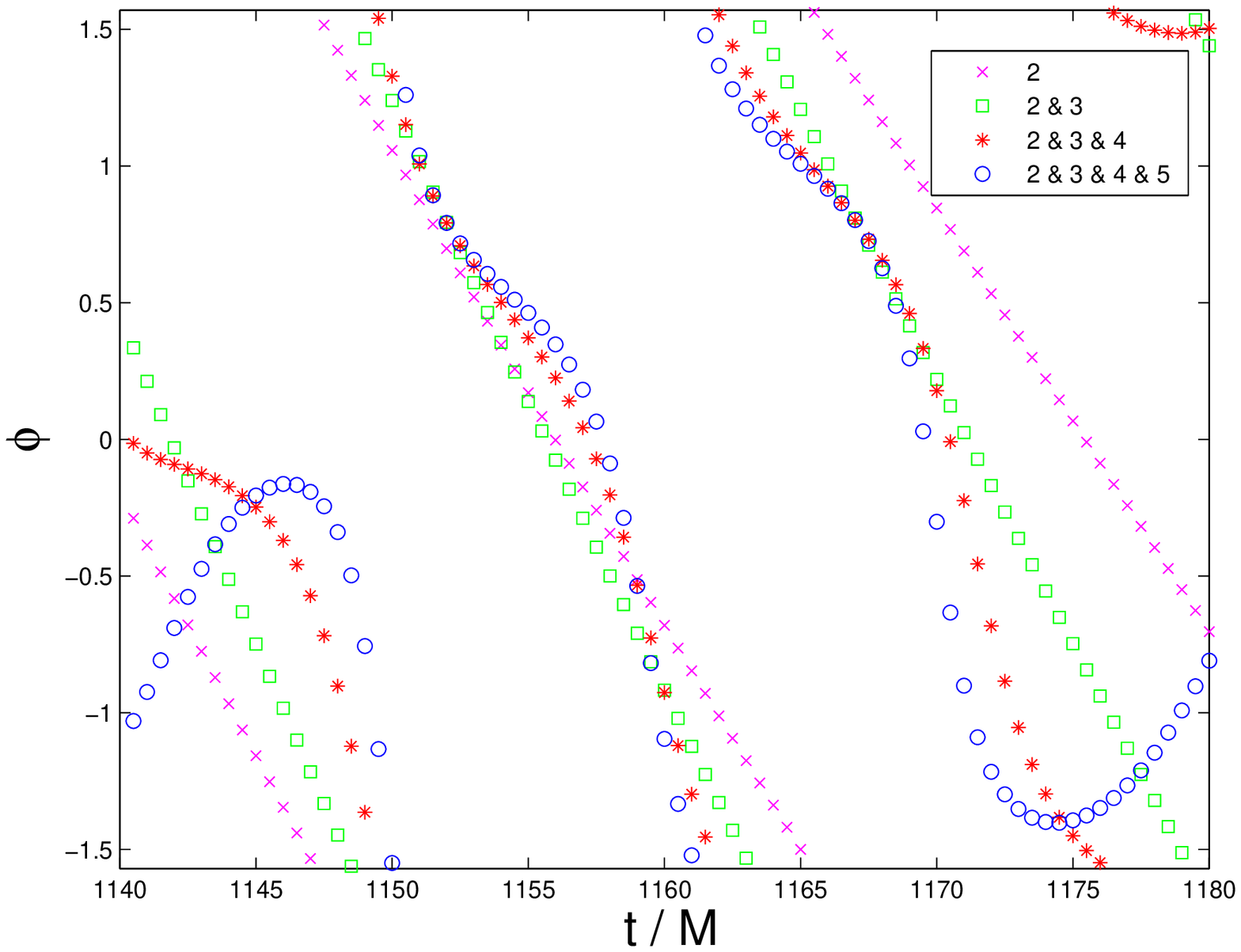}
\caption{The phase of the waveform on the equatorial plane for the same parameters as in Fig.~\ref{fig5re}. Shown are the phases corresponding to $\psi_{m=2}$ ($\times$), $\psi_2+\psi_3$ ($\square$), $\psi_2+\psi_3+\psi_4$ ($*$) and  $\psi_2+\psi_3+\psi_4+\psi_5$ ($\circ$), as a function of $t/M$.  
}
\label{fig5a}
\end{figure}

\section{Summation over multipoles $\ell$ and $k$ modes}\label{k_mode}

The behavior of the $m$-modes ---that corresponds to the angular frequency $\Omega_{\phi}$--- is common to both time-domain and frequency-domain approaches. In the TD case, however, the study of the behavior of the $m$-modes summarizes all the modes that need to be considered to get the full waveform and the total fluxes radiated. In the FD case, however, one needs to consider also the modes that arise from the three different frequencies of the problem: the $k$-modes, corresponding to the radial angular frequency $\Omega_r$, and the $n$-modes, corresponding to the inclination angle oscillations with angular frequency $\Omega_{\theta}$. (We note that various authors exchange the notation of the latter two frequencies, $k \leftrightarrow n$.) Here, we concentrate on the equatorial plane, and defer discussion on motion outside the equatorial plane to a sequel. 

The FD code that we use is based on the code presented in \cite{glampedakis02}, where more details on the code can be found in addition to descriptions of the tests done to check the code and the sources for errors.  The FD numerical method is based on that of \cite{num1} and solves the Sasaki--Nakamura equation \cite{s-n} using Burlisch--Stoer integration. The accuracy level of the code is at the level of $10^{-6}$--$10^{-4}$.\footnote{Recently, possible inaccuracies for very large $k$ values were noticed. These inaccuracies occur for larger values of $k$ than those used here, and therefore do not affect our results.}

\begin{figure}
\input epsf
\includegraphics[width=11.0cm]{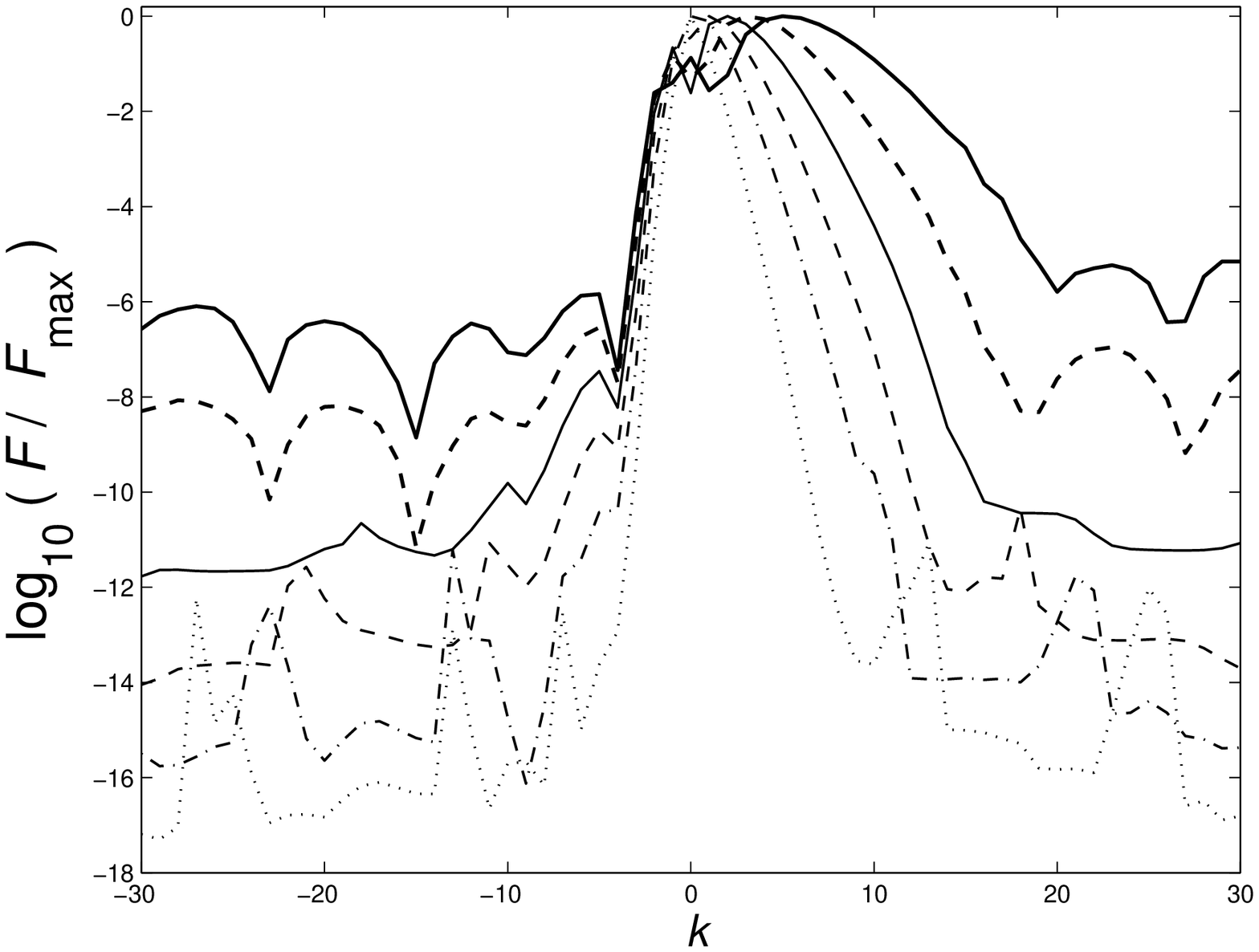}
\caption{The flux in different $k$-modes, after summation over all $\ell$, for $m=2$, for $a/M=0.9$ and $p/M=4.64$: $\epsilon=0.1$ (dotted), $\epsilon=0.2$ (dash-dotted), $\epsilon=0.3$ (thin dashed), $\epsilon=0.4$ (thin solid), $\epsilon=0.5$ (thick dashed), and $\epsilon=0.6$ (thick solid). For each value of the eccentricity, the flux is normalized to the flux at the value of $k$ for which the flux is maximal.}
\label{fig6}
%
\includegraphics[width=11.0cm]{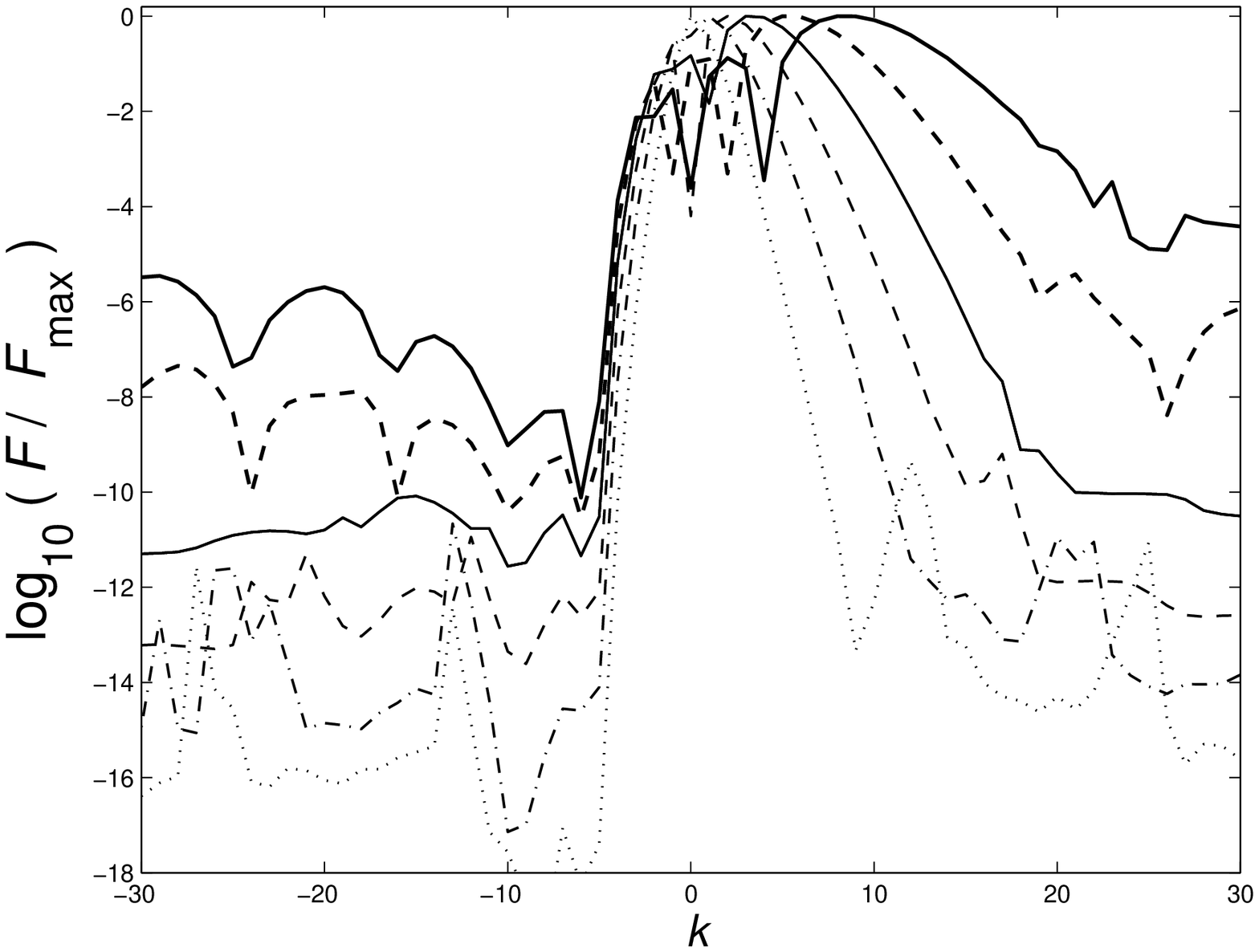}
\caption{Same as Fig.~\ref{fig6} for $m=3$.}
\label{fig6a}
\end{figure}

\begin{figure}
\input epsf
\includegraphics[width=11.0cm]{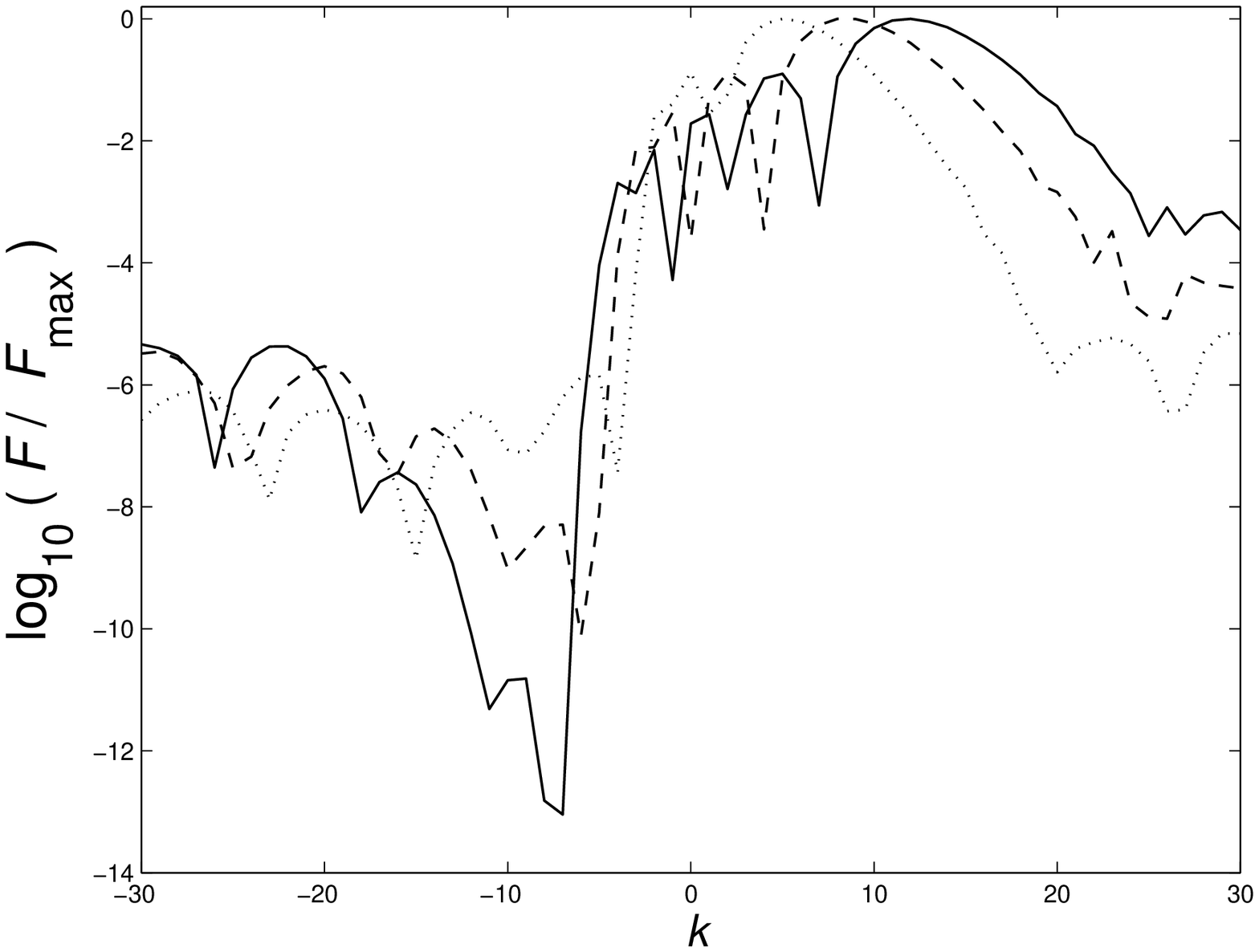}
\caption{Same as Fig.~\ref{fig6}, for $m=2$ (dotted), $m=3$ (dashed), and $m=4$ (solid), for $\epsilon=0.6$. Notice that the first two curves repeat data already presented in Figs.~\ref{fig6} and \ref{fig6a}.}
\label{fig6b}
\end{figure}

First indication to the increase in the number of necessary $k$-modes with increasing eccentricity was shown in \cite{khanna04}. It should be noticed, however, that the results presented in \cite{khanna04} for $\epsilon=0.8$ may be more indication of the failure of the solution of the radial Teukolsky equation in the frequency domain and the need for the Sasaki--Nakamura formulation thereof, than a true behavior of the $k$-modes. Figure \ref{fig6} shows the flux in each $k$ mode for $m=2$, after all $\ell$-modes were summed over, for $a/M=0.9$ and $p/M=4.64$ for different values of the eccentricity. Similar figures were obtained also for other values of $m$. Figures \ref{fig6} and \ref{fig6a} suggest the following  types of behavior: first, the value of $k$ for which the flux is maximal shifts to higher values with the increase in eccentricity of the orbit. Second, with increasing eccentricity, the flux curve broadens (mostly for positive values of $k$, so that the curve becomes less and less symmetric in $k$ for high $\epsilon$). We also find (Fig.~\ref{fig6b}) that the peak of the flux curves become $m$-modal, i.e., at high values of $\epsilon$ the flux curve breaks into a number of peaks equal to the mode $m$. Similar behavior was reported first in \cite{glampedakis02}. 
The peaks background exhibits interesting oscillations, that appear to be independent (or, at the most, weakly dependent) of $m,\epsilon$. The broadening of the flux curves suggests that more $k$-modes are required in order to obtained the total flux. In what follows we quantify this statement and make it precise.

\begin{figure}
\input epsf
\includegraphics[width=11.0cm]{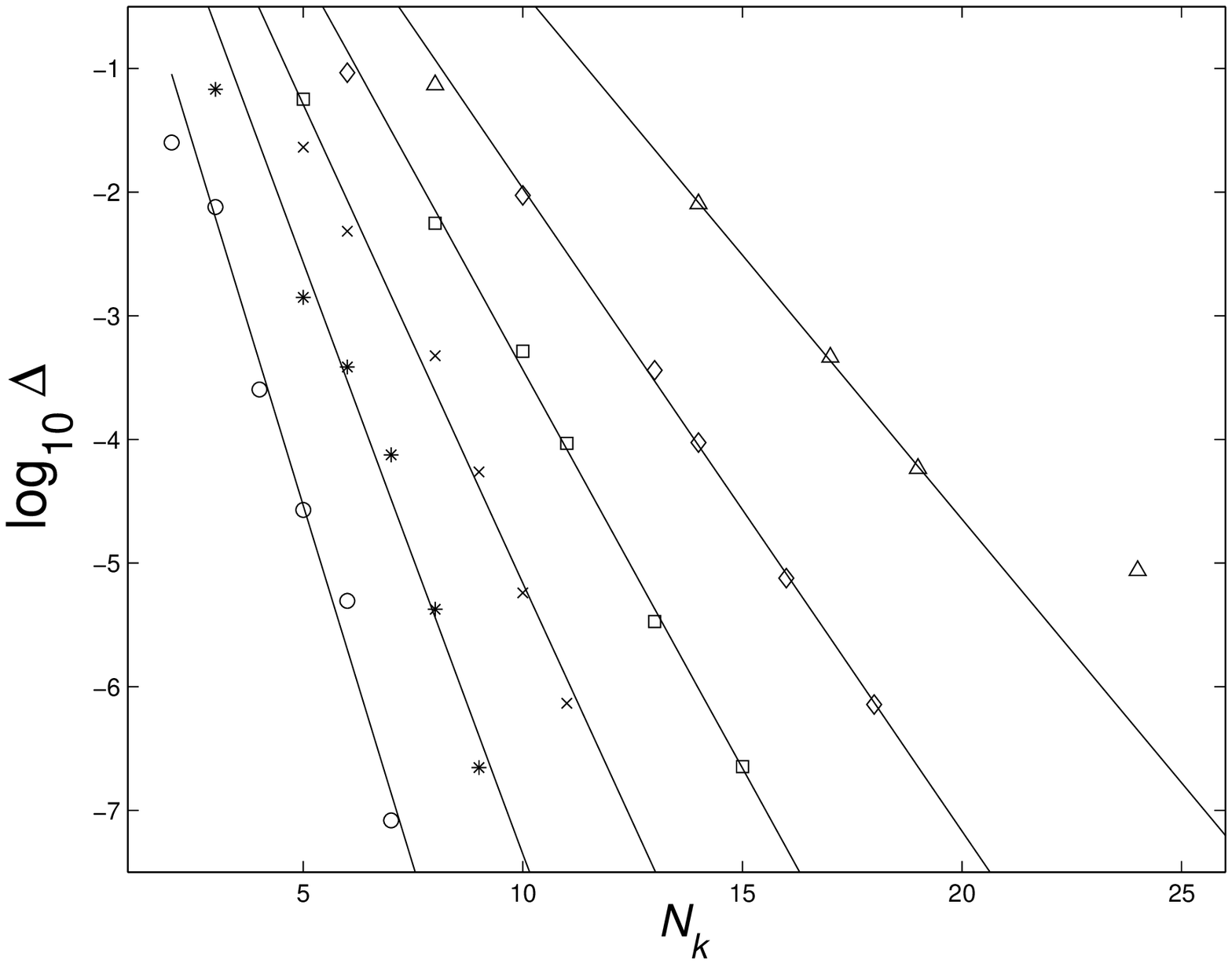}
\caption{The error $\Delta$ in summing over a number $N_k$ of $k$-modes, after summation over all $\ell$, for $m=2$, for $a/M=0.9$ and $p/M=4.64$: $\epsilon=0.1$ ($\circ$) , $\epsilon=0.2$ ($\ast$), $\epsilon=0.3$ ($\times$), $\epsilon=0.4$ ($\square$), $\epsilon=0.5$ ($\diamond$), and $\epsilon=0.6$ ($\triangle$). For each value of the eccentricity, the solid line describes a fitted line as follows, corresponding to increasing eccentricity: 
$\log_{10}\Delta=1.2796-1.1630\,N_k$ ($R^2=0.9810$), 
$\log_{10}\Delta=2.2131-0.9567\,N_k$ ($R^2=0.9691$), 
$\log_{10}\Delta=2.5781-0.7740\,N_k$ ($R^2=0.9774$), 
$\log_{10}\Delta=3.0085-0.6445\,N_k$ ($R^2=0.9962$), 
$\log_{10}\Delta=3.2273-0.5197\,N_k$ ($R^2=0.9987$), 
$\log_{10}\Delta=3.8898-0.4267\,N_k$ ($R^2=0.9383$). For each fitted curve we include the square of the correlation coefficient $R^2$}
\label{fig7}
\end{figure}

\begin{figure}
\input epsf
\includegraphics[width=11.0cm]{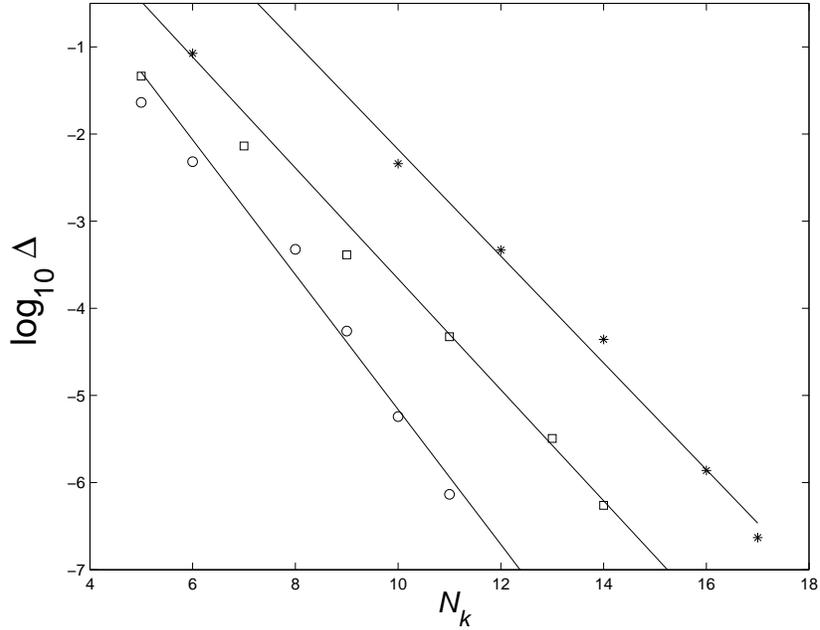}
\caption{The error $\Delta$ in summing over a number $N_k$ of $k$-modes, after summation over all $\ell$, for $a/M=0.9$, $p/M=4.64$, and $\epsilon=0.3$: $m=2$ ($\circ$), $m=3$ ($\square$), and $m=4$ ($\ast$). For each value of the eccentricity, the solid line describes a fitted line as follows, corresponding to increasing value of $m$: 
$\log_{10}\Delta=2.5781-0.7740\,N_k$ ($R^2=0.9774$), 
$\log_{10}\Delta=2.7038-0.6366\,N_k$ ($R^2=0.9950$), 
$\log_{10}\Delta=3.9493-0.6126\,N_k$ ($R^2=0.9893$). 
 For each fitted curve we include the square of the correlation coefficient $R^2$}
\label{fig8}
\end{figure}

\begin{figure}
\input epsf
\includegraphics[width=11.0cm]{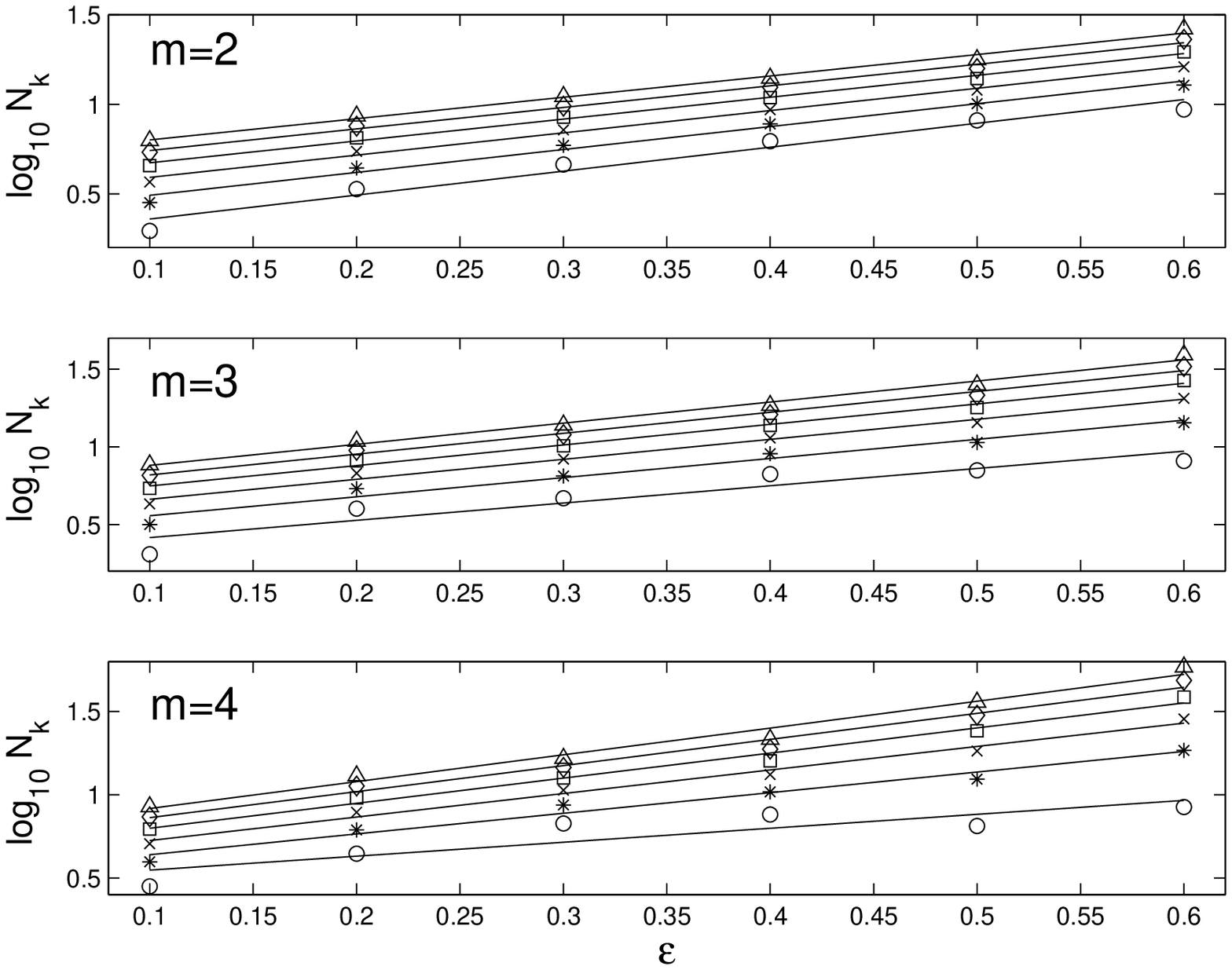}
\caption{The number $N_k$ of $k$-modes required for a given accuracy as a function of the 
eccentricity $\epsilon$. Shown are 6 levels of accuracy, and their corresponding contours: 
$10^{-1}$ ($\circ$), $10^{-2}$ ($\ast$), $10^{-3}$ ($\times$), $10^{-4}$ ($\square$), $10^{-5}$ 
($\diamond$), and $10^{-6}$ ($\triangle$), for three values of the mode number $m=2,3,4$. 
Upper panel ($m=2$): the solid lines describe a fitted line as follows, corresponding the increasing accuracy: 
$\log_{10}N_k=0.2253+1.3355\,\epsilon$ ($R^2=0.9642$), 
$\log_{10}N_k=0.3632+1.2783\,\epsilon$ ($R^2=0.9872$),
$\log_{10}N_k=0.4671+1.2442\,\epsilon$ ($R^2=0.9945$), 
$\log_{10}N_k=0.5506+1.2215\,\epsilon$ ($R^2=0.9959$),
$\log_{10}N_k=0.6206+1.2053\,\epsilon$ ($R^2=0.9948$), 
$\log_{10}N_k=0.6808+1.1930\,\epsilon$ ($R^2=0.9928$).
Middle panel ($m=3$): the solid lines describe a fitted line as follows, corresponding the increasing accuracy: 
$\log_{10}N_k=0.3040+1.1118\,\epsilon$ ($R^2=0.8860$), 
$\log_{10}N_k=0.4321+1.2321\,\epsilon$ ($R^2=0.9727$),
$\log_{10}N_k=0.5329+1.2886\,\epsilon$ ($R^2=0.9900$), 
$\log_{10}N_k=0.6151+1.3216\,\epsilon$ ($R^2=0.9934$),
$\log_{10}N_k=0.6844+1.3433\,\epsilon$ ($R^2=0.9932$), 
$\log_{10}N_k=0.7442+1.1588\,\epsilon$ ($R^2=0.9919$).
Lower panel ($m=4$): the solid lines describe a fitted line as follows, corresponding the increasing accuracy: 
$\log_{10}N_k=0.4636+0.8385\,\epsilon$ ($R^2=0.7739$), 
$\log_{10}N_k=0.5159+1.2410\,\epsilon$ ($R^2=0.9747$),
$\log_{10}N_k=0.5841+1.4111\,\epsilon$ ($R^2=0.9894$), 
$\log_{10}N_k=0.6478+1.5066\,\epsilon$ ($R^2=0.9884$),
$\log_{10}N_k=0.7050+1.5681\,\epsilon$ ($R^2=0.9853$), 
$\log_{10}N_k=0.7564+1.6111\,\epsilon$ ($R^2=0.9821$).
}
\label{fig9}
%
\includegraphics[width=11.0cm]{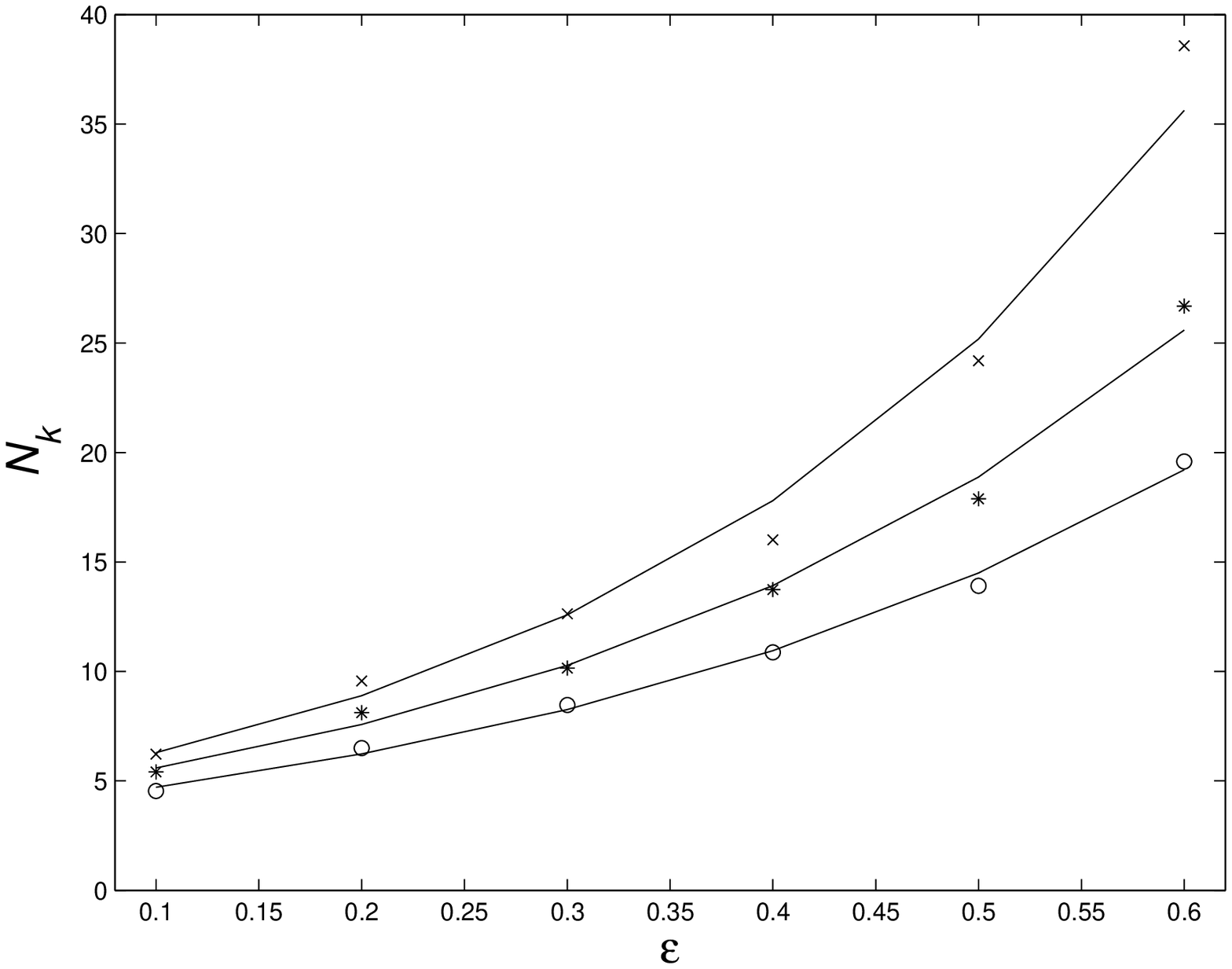}
\caption{The number $N_k$ of $k$-modes required for an accuracy of $10^{-4}$ as a function of the 
eccentricity $\epsilon$. Shown are 3 values of $m$, and their corresponding contours, based on a fit to an exponential:
$m=2$ ($\circ$), $m=3$ ($\ast$), and $m=4$ ($\times$). 
}
\label{fig10}
\end{figure}

The number of $k$-modes required at a fixed value of the eccentricity $\epsilon$ to obtain the total flux to a desired accuracy increases logarithmically with the accuracy. This behavior is described in Fig.~\ref{fig7} that plots the error involved in the inclusion of the $N_k$ modes (of the greatest partial fluxes) as a function of $N_k$ for various values of the eccentricity $\epsilon$. In all cases this behavior is exponential. Notice, however, that the slope of the fitted curve becomes steeper with increasing eccentricity. This behavior suggests that as the eccentricity increases, the rate at which the number of $k$-modes that are required in order to obtain the same level of accuracy increases. This rate also increases with the mode number $m$ (Fig.~\ref{fig8}). Notice, from Fig.~\ref{fig6b}, that also the number of {\em negative} $k$-modes increases with the mode number $m$ (for fixed eccentricity), although not as fast as the number of positive $k$-modes. In Fig.~\ref{fig9} we show the number $N_k$ of $k$-modes required for a given accuracy as a function of the eccentricity $\epsilon$, for different values of the mode number $m$. The increase in $N_k$ is exponential in the eccentricity $\epsilon$. For a given accuracy level, the exponential increase with $\epsilon$ becomes steeper as the mode number $m$ increases, as is shown in Fig.~\ref{fig10}. 

\section{Modes computation time}\label{comp_time}

The computation time of a single $k$ mode is a function of $k$. To make a prediction of the $k$ dependence of the computation time for a $k$ mode we first write the radial Teukolsky equation 
\begin{equation}\label{radial_teukolsky}
\Delta^2\,\frac{\,d}{\,dr}\,\left(\Delta^{-1}\,\frac{\,dR_{\ell m \omega}}{\,dr}\right)-V(r)\,R_{\ell m \omega}(r)=-{\cal T}_{\ell m \omega}(r)
\end{equation}
where the potential $V(r)$ is a complex valued function of $r$ and the angular frequency $\omega$ \cite{hughes-2000}. The source term ${\cal T}_{\ell m \omega}(r)$ is a certain known function of the stress energy of the particle and its trajectory, and of the background geometry. As we are interested here only in the qualitative features of the solutions for Eq.~(\ref{radial_teukolsky}), the details of the source are unimportant for us here. We next focus attention on the far field limit of Eq.~(\ref{radial_teukolsky}). To gain a qualitative understanding of the structure of the solution, we next consider only the asymptotic solution as $r\to\infty$. In that limit, $R(r\to\infty)\sim r^3\,e^{i\omega r_*}$ (corresponding to an outgoing solution), where $r_*$ is the usual Kerr spacetime tortoise coordinate defined by $\,dr_*/\,dr=(r^2+a^2)/\Delta$. The solution is an oscillatory solution in $r$ with angular frequency $\omega$. The typical length scale over which the solution is oscillating is therefore $\lambda\sim 2\pi/\omega$: the greater $\omega$, the shorter the distance over which the solution oscillates.

\begin{figure}
\input epsf
\includegraphics[width=11.0cm]{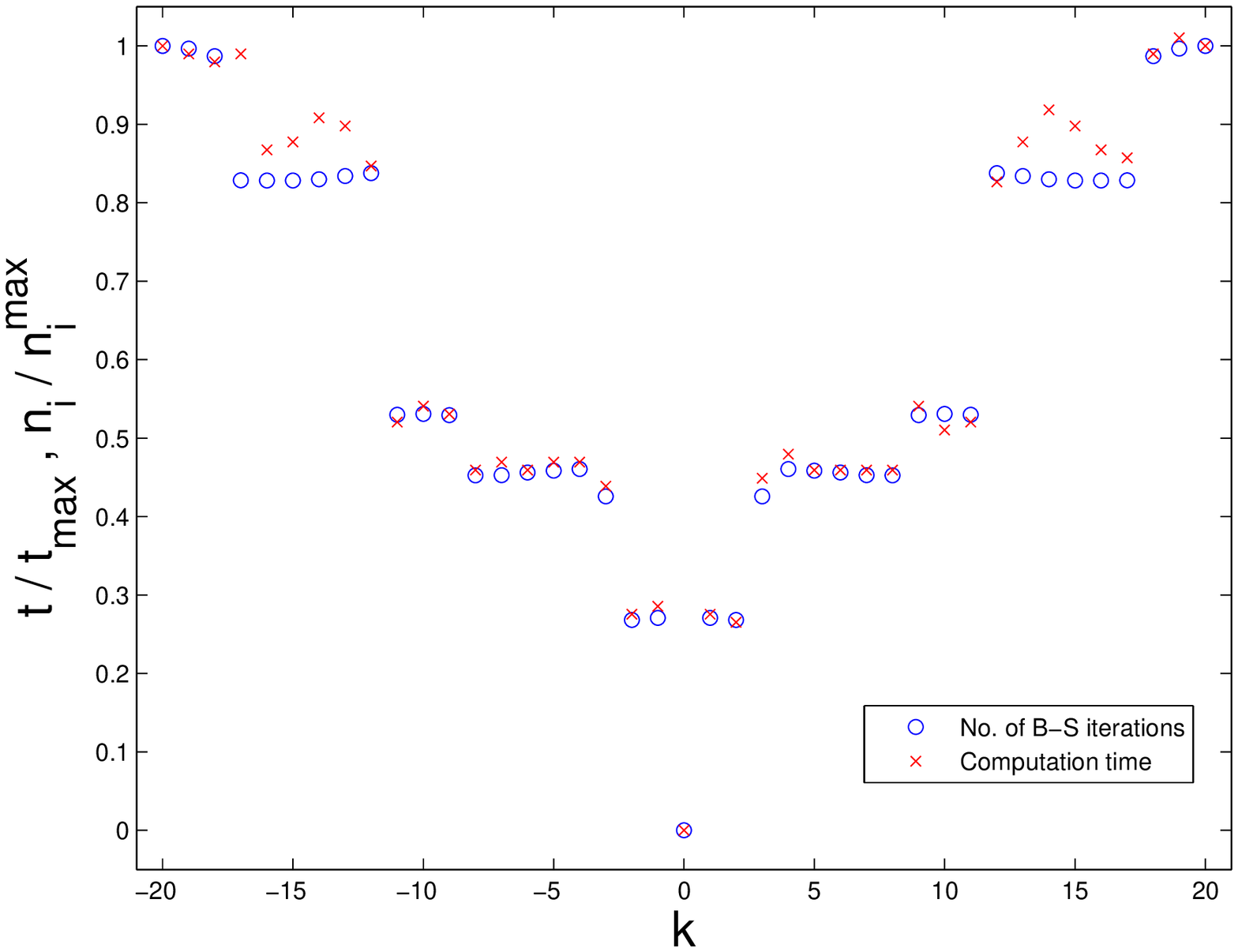}
\caption{The dependence of the computation time of a $k$ mode on $k$: the computation time $t$ (normalized by the maximal computation time is shown in $\times$,  and the corresponding normalized number of iterations $n_i$ that the Burlisch--Stoer engine does is shown in $\circ$. The data are taken for $m=0$, and for $p/M=4.64$ and $\epsilon=0.4$.
}
\label{fig11}
\includegraphics[width=11.0cm]{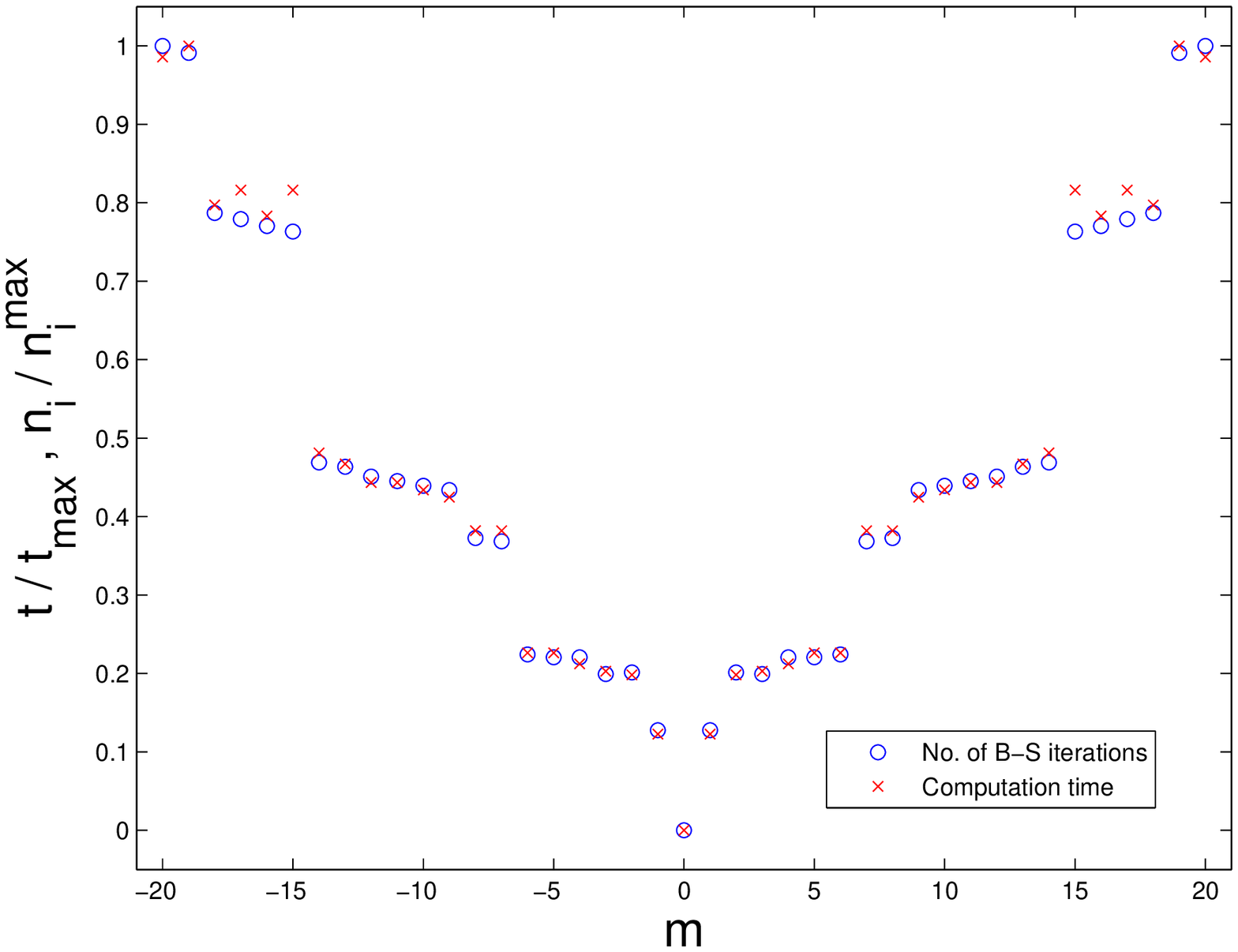}
\caption{Same as Fig.~\ref{fig11} for the $m$ modes. Here, $k=0$ and $\ell=20$. 
}
\label{fig12}
\end{figure}

The method we use to solve the frequency-domain equation is by a Burlisch--Stoer algorithm \cite{num_rec}, that successively divides a single step into many substeps, until a (polynomial or rational) interpolation is accurate enough. If the original step happens to be too large (i.e., after a certain pre-determined substep is calculated and the required accuracy is still not obtained), the original step will be bisected, and the process proceeds. Therefore, the greater $\omega$ and the shorter the oscillation length scale, the more divisions of the initial step are required, and the greater the number of substeps used to find the solution.  
As the angular frequency $\omega_{mk}=m\Omega_{\varphi}+k\Omega_r$, it is clear that the larger $k$ (for a fixed value of $m$), the shorter the distance scale $\lambda$. We therefore expect the computation time of a single $k$ mode to increase with $k$. In fact, the Burlisch--Stoer algorithm suggests that as $\omega_{mk}$ increases, the more substeps are needed, until for some value of $k$ dividing the step into substeps is no longer sufficient, and the step is bisected. Then, division into substeps is again sufficient, until another $k$ value is obtained for which the step is bisected. Each time the step is bisected, the total number of substeps computed jumps discontinuously, so that we expect the $k$ mode computation time to increase as a staircase function. In Fig.~\ref{fig11} we show the computation time of an individual $k$ mode as a function of $k$, and also the number of Burlisch--Stoer iterations done. They both behave as expected. The preceding discussion suggests that a  similar behavior of the computation time is found also as a function of $m$. This is indeed seen in Fig.~\ref{fig12}.

As can be seen from Figs.~\ref{fig11} and \ref{fig12} the discontinuous jumps in the mode computation times occur at different values of $m,k$. Indeed, our discussion about explains this behavior: jumps occur when the angular frequencies $\omega_{mk}$ arrive at some threshold values, corresponding to threshold wavelengths. As these threshold values should be about the same, we expect jumps to occur for $m_j,k_j$ satisfying $k_j\Omega_r\sim m_j\Omega_{\varphi}$. Notice that $k_j$ ($m_j$) is defined here for vanishing $m$ ($k$). 
For the data presented here, for $\epsilon=0.4$ we find $k_j/m_j\approx 1.33$, while $\Omega_{\varphi}/\Omega_r\approx 1.52$; for  $\epsilon=0.5$ we find $k_j/m_j\approx 1.40$, while $\Omega_{\varphi}/\Omega_r\approx 1.47$; and for $\epsilon=0.6$ we find $k_j/m_j\approx 1.44$, while $\Omega_{\varphi}/\Omega_r\approx 1.39$, so that the difference between $k_j/m_j$ and $\Omega_{\varphi}/\Omega_r$   is at order 10\%. We argue that these results support our interpretation of the discontinuous jumps in the computation time.

\begin{figure}
\input epsf
\includegraphics[width=11.0cm]{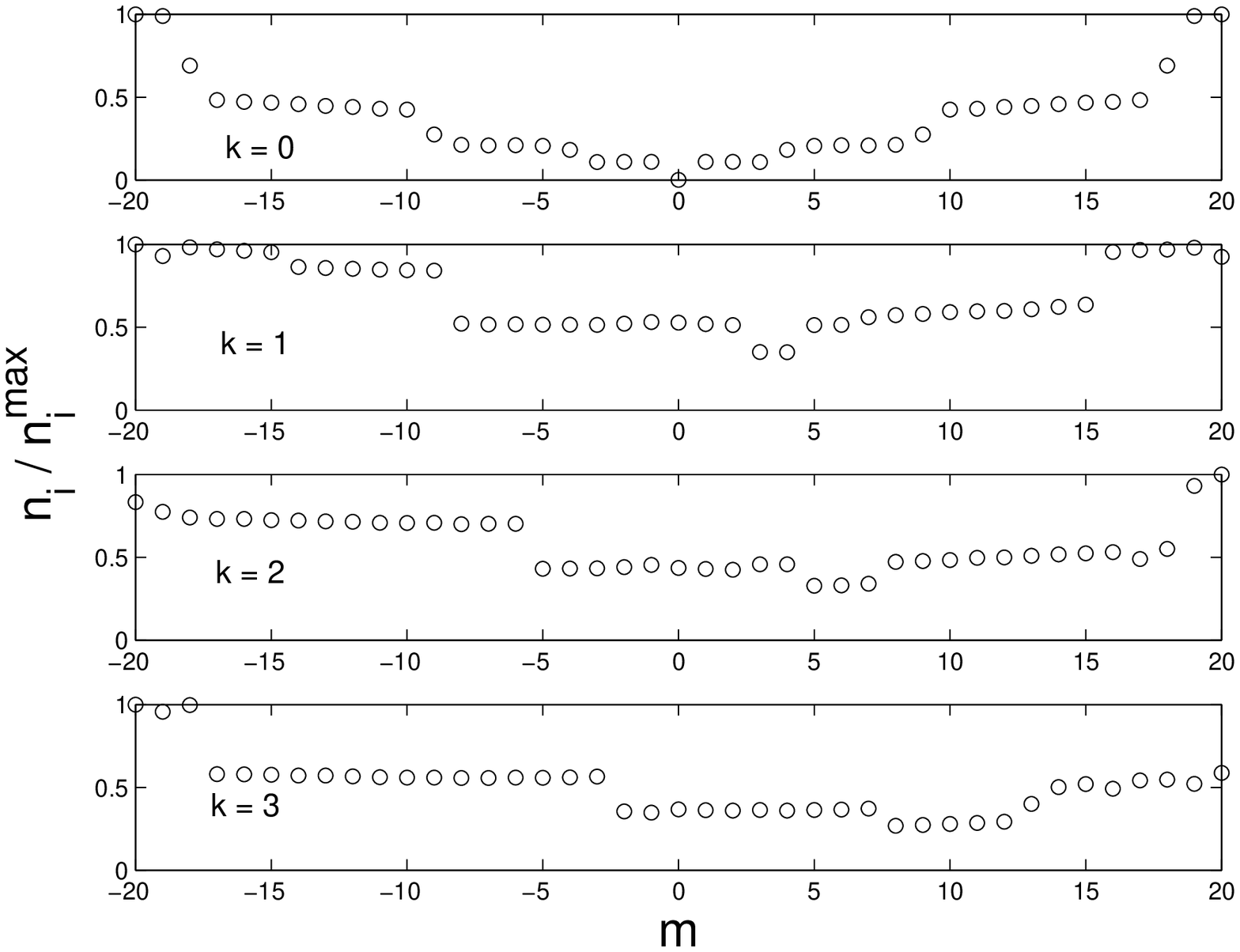}
\caption{Number of iterations of the Burlisch--Stoer solver as a function of $m$ for fixed values of $k$, for the same orbital parameters as above, for $\epsilon=0.1$. 
}
\label{fig13}
\end{figure}

Our discussion suggests that for a fixed value of $k$ ($m$) and varying $m$ ($k$), there are values of the latter for which the computational time drops. These are the values that approximately satisfy $\omega_{mk}\lesssim \Omega_{\varphi}\; ,\; \Omega_{r}$. (Recall that $k,m$ can be either positive or negative.) That is, while fixing either $k$ or $m$ and varying the other, 
we expect to find a drop in the computation time (or, equivalently, in the number of Burlisch--Stoer iterations.) Indeed, we find this behavior, as is shown in Fig.~\ref{fig13}. Notice, that for $k=0$ the drop in computational time is found for $m=0$, and as $k$ increases, so does the value of $m$ for which the drop is found. We also comment that this drop is broadened with increasing values of $k$.

The dependence of the computational time on the mode number $k$ is therefore rather intricate. It can be approximated as follows: For each value of the eccentricity $\epsilon$ we take in practice the  computational time of a single mode $k$ to be a linear function of $k$, that fits the computational data. This approximation underestimates the computation time for some $k$ modes, specifically those immediately following a discontinuous jump in the computation time, and overestimates the computation time for $k$ modes just before a jump. However, as we are mostly interested in the total computation time of the sum over all modes, this approximation may be quite reasonable for the sum over all modes. In this approximation the sum over $k$ modes up to some $k_{\rm max}$ is a quadratic function of $k_{\rm max}$. Notice, that we can use Fig.~\ref{fig7} and \ref{fig8} to find how many $k$ modes we need to sum over to obtain the required accuracy level. 

\begin{figure}
\input epsf
\includegraphics[width=11.0cm]{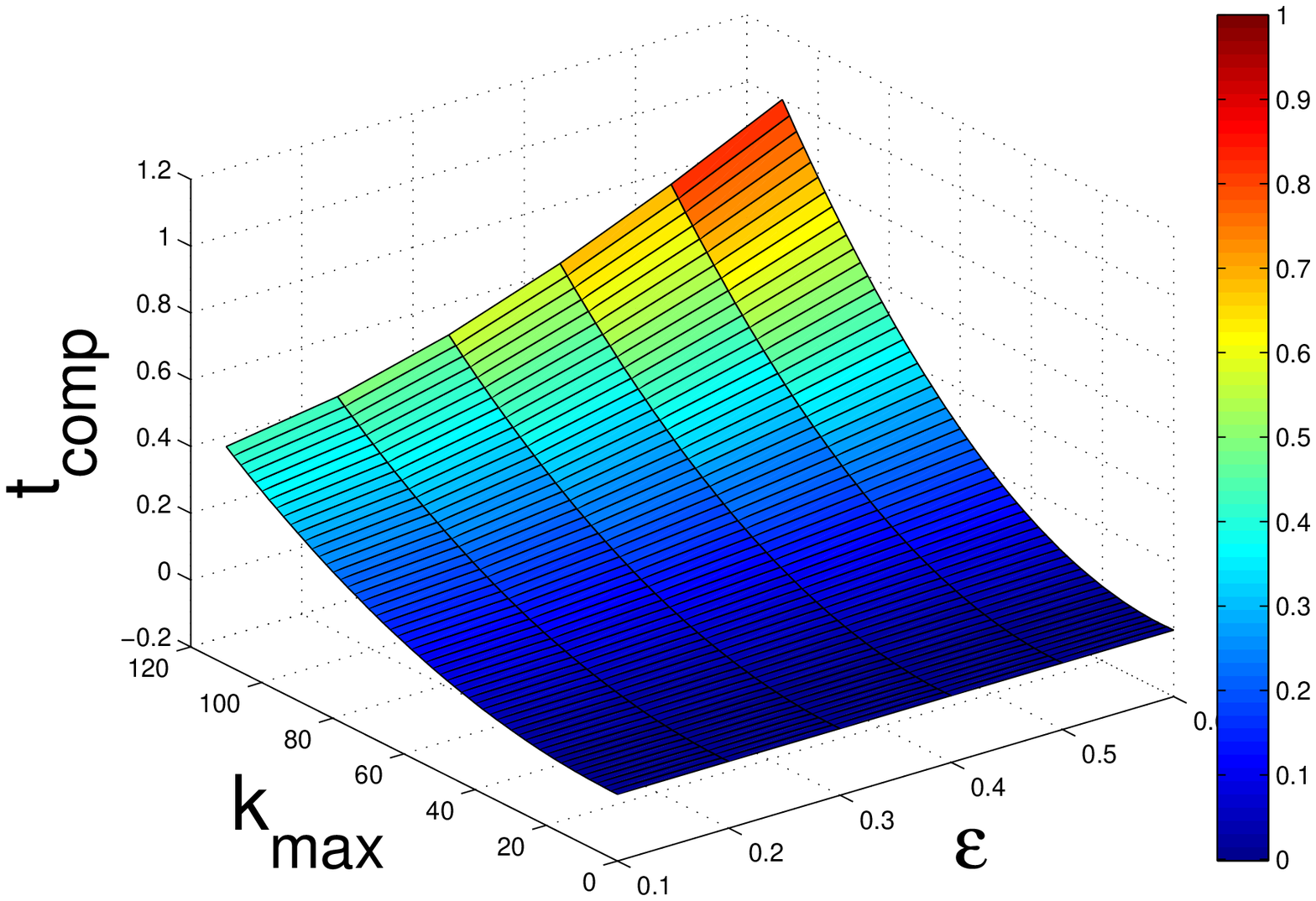}
\caption{The total computation time of the sum over $k$ modes from $k=0$ to $k=k_{\rm max}$ as a function of the eccentricity $\epsilon$. The computation time is presented in units of the maximal computation time in the chosen range of parameters. 
}
\label{fig14}
\end{figure}

\section{Comparison of total computation time of TD and FD calculations}\label{compare}

Finally, we put together all the previous results, to compare the actual total calculation time of TD and FD codes. We choose an orbit that gives no special preference for one approach over the other, i.e., a moderate value for the eccentricity. In practice, we take $\epsilon=0.5$ and $p=4.64M$. 

For the FD calculation we first set the desired level of accuracy, which in practice we choose here to be $10^{-3}$, and then use Fig.~\ref{fig4} (or the data in Table \ref{table5}) to determine the number of $m$ modes one needs to sum over to guarantee the desired accuracy level. The needed number of $m$ modes is found in practice to be 7, for $m=2,3,\cdots, 9$.  

\begin{figure}
\input epsf
\includegraphics[width=11.0cm]{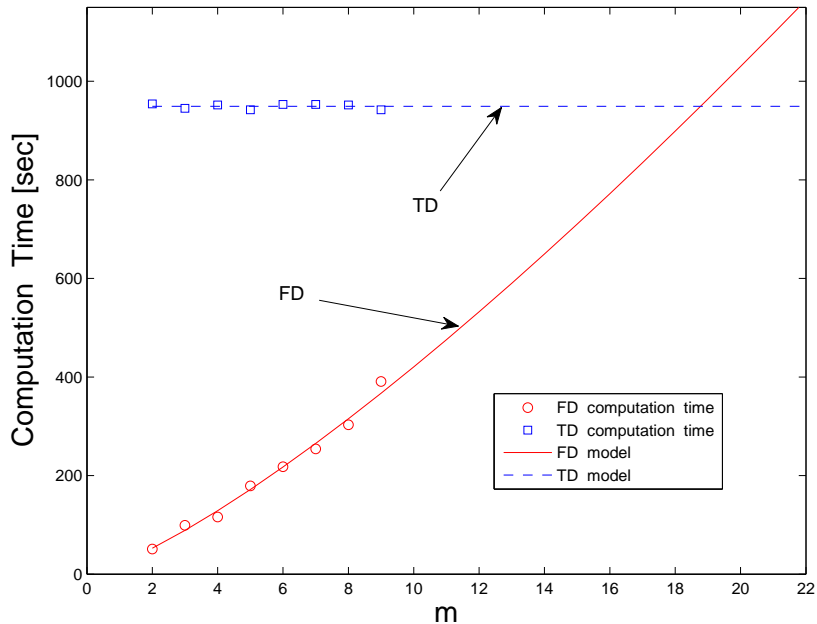}
\caption{Comparison of computation time for FD and TD for an orbit with $p/M=4.64$ and $\epsilon=0.5$ as a function of the mode number $m$. The  circles ($\circ$) denote FD computation time, and the squares ($\square$) show TD computation times. Both sets of runs were done on the same computer, in this case a single processor of a dual 2.7GHz PowerPC G5. The red curve is the fitted curve $t=21.471\,m^{1.2919}\;{\rm sec}$, that fits the numerical data with a squared correlation coefficient of $R^2=0.9895$. For both cases the accuracy level is set at $10^{-3}$.
}
\label{fig15}
\end{figure}

For each individual $m$ mode we then determine the number of $k$ modes necessary to obtain the desired accuracy level. This determination is done with data as in Figs.~\ref{fig9} and \ref{fig10}. The values of the $k$ modes we used in practice are as follows: for $m=2$ we used $-2\leqslant k\leqslant 10$, for $m=3$ we used $-3\leqslant k\leqslant 13$, for $m=4$ we used $-3\leqslant k\leqslant 16$, for $m=5$ we used $-2\leqslant k\leqslant 21$, for $m=6$ we used $-1\leqslant k\leqslant 23$, for $m=7$ we used $-2\leqslant k\leqslant 25$, for $m=8$ we used $-1\leqslant k\leqslant 28$, and for $m=9$ we used $-1\leqslant k\leqslant 33$. In all cases we computed $2\leqslant \ell\leqslant 5$. The data are summarized in Table \ref{table6} and in Fig.~\ref{fig15}. The FD calculations were optimized for most efficient calculation for the desired accuracy level. Improvements to the computation time can be achieved, but only at the level of the code, e.g., making the Burlisch--Stoer algorithm more efficient. We believe that while such improvements can be made, their effect would be moderate. In this sense, the FD curve in Fig.~\ref{fig15} represents a lower bound on the FD calculation time. Increasing the eccentricity of the orbit $\epsilon$ would result in the FD curve of Fig.~\ref{fig15} moving up in the figure (increase in the calculation time of each $m$ mode, because of increasing number of required number of $k$ modes and increase in the computation time of individual $k$ modes) and also having a faster growth rate.

\begin{table}[h]
 \caption{Number of $k$ modes for each $m$ modes in the FD calculation, and the corresponding computation time in both the FD and TD calculations. The orbital parameters are $\epsilon=0.5$ and $p/M=4.64$ and $a/M=0.9$.}
  \centering
     \begin{tabular}{|c||c|c|c||c|} \hline
   {\bf m mode} & Range of $k$ modes & Total $k$ modes &FD time (sec) & TD time (sec)
   \cr \hline \hline
   2 & $-2\leqslant k\leqslant 10$ & 13 & 51 & 954   \cr \hline
   3 & $-3\leqslant k\leqslant 13$ & 17& 99 & 945  \cr \hline
   4 & $-3\leqslant k\leqslant 16$ & 20 & 116 & 952  \cr \hline
   5 & $-2\leqslant k\leqslant 21$ & 24 & 179 & 942  \cr \hline
   6 & $-1\leqslant k\leqslant 23$ & 25 & 218 & 953   \cr \hline
   7 & $-2\leqslant k\leqslant 25$ & 28 & 254 & 953  \cr \hline
   8 & $-1\leqslant k\leqslant 28$ & 30 & 303 & 952  \cr \hline
   9 & $-1\leqslant k\leqslant 33$ & 35 & 391 & 942  \cr \hline \hline
   \end{tabular}
\label{table6}
\end{table}

For the TD calculations we used the 2+1D code of \cite{burko-_khanna_07}, with radial resolution of $\,\Delta r=M/20$ (and temporal resolution of $\,\Delta t=\,\Delta r/2$), and angular resolution of $\,\Delta\theta=\pi/32$. We placed the inner and outer boundaries at $r_*/M=-50$ and at $r_*/M=350$, respectively. We approximate the scattering problem on the boundaries by assuming Sommerfeld boundary conditions, i.e., no incoming radiation from outside the boundaries. One may of course make this approximation better by pushing the outer boundary outwards (and the inner boundary inwards). Reflections from the boundaries then do not contaminate the center of the computational domain (at $r_*/M=150$) until $t/M=400$, and we integrate in time until then. In practice, we wait at each evaluation point until the initial spurious waves (that result from imprecise initial data that correspond to a particle suddenly appearing at $t/M=0$), and then average the flux over a full period of the orbit. The angular period is $T_{\phi}=2\pi\,M^{-1/2}\,(r^{3/2}+aM^{1/2})$, where $r$ is the semimajor axis. For our choice of parameters, $T_{\phi}\approx 102.4M$. The radial period is found to be $T_r\approx 162.5M$, and it is the latter (radial) period we use for our analysis. 

We extract the (radial period averaged) flux at a number of extraction distances, in practice at $r_*/M=30,40,\cdots,100$, and then fit the (finite extraction distance) fluxes to an inverse-square function as in Eq.~(\ref{ansatz}). This allows us to achieve fluxes that agree with the FD fluxes to $10^{-3}$, while not having to integrate to very late times (and commensurately increase also the spatial computational domain). Notably, we can obtain higher accuracy than that reported in \cite{burko-_khanna_07} while integrating to a shorter time because we make use of the finite time corrections to the energy flux. We discuss this method in detail in the Appendix. 

The TD computational time is presented in Table \ref{table6}. It can be made more efficient using code improvements, and data analysis improvements. First, we used gaussian source modeling as in \cite{burko-_khanna_07} , and the discrete $\delta$ approach  may improve efficiency considerably. In fact, Ref.~ \cite{pranesh} has argued for a full order of magnitude reduction in computation time. While this statement may be an overestimate of the numerical capabilities ---especially for generic (i.e., eccentric or inclined) orbits--- it is certainly possible to improve the efficiency of the TD code by changing the method of calculation of the source term. (The discrete delta approach has also other advantages over the gaussian model. See \cite{pranesh} for more detail.) An important property of the TD code is that it is very naturally parallelizable. Certainly the FD computation is also parallelizable, most naturally by having each $k$ mode computed on a different processor. In such a case, the total FD computation time is controlled by the $k$ mode that takes the longest to compute, i.e., the largest $k$ (see Fig.~\ref{fig11}). Another improvement of efficiency of the FD code may come from improvements of the Burlisch--Stoer engine, specifically the solution for $\log r$ instead of $r$. It is currently unknown how the computational efficiency and numerical accuracy would be affected by such a change. 

One factor that limits a possible reduction of the TD computation time is the need to average over an orbital (radial) period to find the average flux per period. Even for the strong field orbit considered here, with $p/M=4.64$ and $\epsilon=0.5$, the radial period is $T_r=162.5M$, which puts a lower bound on how short the computational integration can be. One could perhaps shorten the computation time if only half an orbital period is computed, but because the phase of the orbit at the end of the spurious wave epoch is arbitrary, on the average at least three quarters of an orbit need to be computed. However, we were interested here in comparison of the fluxes to infinity. If one is interested in the wave function itself, shorter computation times can be used, thus saving much of the TD computation time. No equivalent save in time can be achieved with the FD computation. Reduction of the TD computation time would result in the dashed line in Fig.~\ref{fig15} moving lower, thus making the computational time of TD and FD comparable already for lower $m$ values. Most importantly, when higher eccentricity orbits are considered, the TD computation time remains unchanged, whereas the FD computation time increases considerably. More specifically, not only is the solid curve in Fig.~\ref{fig15} shifted up, it also becomes steeper (i.e., its rate of growth increases), so that the TD and FD computation times become comparable for lower $m$ values. In addition, the TD calculation does not require any extra computation to find the waveform, as it is already available. In fact, an extra computation was needed to find the fluxes. On the other hand, the FD calculation requires an extra computation to produce the waveform, which we have not done here. Therefore, the greater efficiency of FD over TD for the modes shown here is much less impressive when the computation time for waveforms is of  interest. 

We therefore suggest that at high eccentricity orbits the computational efficiency of the TD and the FD approaches is comparable for high $m$ values. We believe that the question of the total (sum over $m$) computation time is not necessarily a very important one, and suggest that for actual computations the low $m$ modes are calculated using the FD approach, and the high $m$ modes are calculated using the TD approach. The determination of which $m$ values are low and which are high depends of course on the parameters of the system, and the accuracy level of the computation. Using a single method, however, could still possibly make TD more efficient than FD for very high values of the eccentricity.

In the analysis of this paper we have focused on equatorial orbits. Such orbit simplify the problem not just computationally: the number of dimensions in the FD parameter space is significantly lower than that for generic orbits. While the analysis of equatorial orbits presented here is non-trivial, the added complications of handling generic orbits warrant separate treatment. When generic orbits are considered, finding the number of $\ell$ modes required to achieve a set accuracy level with the FD approach becomes a non-trivial problem, as is the question of finding the computational time of each $\ell$ mode as a function of the mode number. Notably, the TD calculations remain unchanged, as is the associated computation time. We therefore expect the FD computation time to increase significantly with the transition from equatorial to inclined and then generic orbits, an increase that has no counterpart in the TD case. To do such an analysis, one needs to check the behavior of all combinations of the modes $k,\ell,m$, find the minimal number of modes required for a set accuracy level, and then compute those modes. The TD calculation requires that the same number of $m$ modes are computed, and this computation time depends only on the duration of the computation, the size of the computational domain, and of course the grid resolution. Modest improvements to the TD computation time may be achieved by improving the efficiency of the source calculations (e.g., the discrete delta approach), but they are not expected to be as dramatic as previously suggested \cite{pranesh}. We expect the threshold eccentricity above  which the TD approach is more computationally efficient than the FD approach for some of the needed $m$ modes to be moderate--high, specifically in the range of $\epsilon\sim 0.6$--$0.7$. 

\section*{Acknowledgments} 

The authors are indebted to Kostas Glampedakis for discussions and  for making his FD code available to us. 
LMB was supported in part by  NASA/GSFC grant No.~NCC5--580, NASA/SSC grant No.~NNX07AL52A, and by a minigrant from the Office of the Vice President for Research at UAHuntsville. GK is grateful for research support from the University of Massachusetts (Dartmouth) -- College of Engineering, the HPC Consortium and Sony Corporation of America (SCEA). 
DJL was supported by a Research Experiences for Undergraduates in Science and Engineering fellowship, sponsored by the 
Alabama Space Grant Consortium under Contract NNG05GE80H and by the UAHuntsville President's Office.

\begin{appendix}
\section{Notes on wave--extraction distance for the time domain code}\label{appendixA}
The peeling theorem describes how the various components of the gravitational field of a bound gravitational system behave as the observer moves away from the source (see, e.g., \cite{stephani,stewart}). Specifically, in the wave zone of a radiating system, the $\Psi_4$ component of the Weyl curvature--- which describes in the wave zone the outgoing transverse radiative degrees of freedom ---typically drops off as $\Psi_4\sim r^{-1}$, where $r$ is the distance from the  radiating system. 
The peeling theorem also gives the drop off rates for the remaining Weyl scalars, and also for spin coefficients and other Newman--Penrose quantities. In addition to the leading order decay rate, the peeling theorem has been extended also to include correction terms in $r^{-1}$. Specifically, 
Newman and Unti  showed that \cite{newman_unti_62}
$$\Psi_4=\frac{\Psi_4^0}{r}-\frac{2\alpha^0\Psi_3^0+{\bar \xi}^{0k}\Psi^0_{3\; ,k}}{r^{2}}+O(r^{-3})\, ,$$
where $\alpha$ is a spin coefficient and $\xi^i$ is an ``integration constant" from the radial equations. A superscript ``$0$" means the coefficient of the slowest decaying term in an expansion in $r^{-1}$. 
As the flux of energy in gravitational waves at infinity ${\cal F}\sim\lim_{r\to\infty} r^2\,\Psi_4^2\sim (\Psi_4^0)^2$, $\Psi_4^0$ is directly related to the flux at infinity. 

Notably, the leading order corrections in $r^{-1}$ to $\Psi_4$ are proportional to the {\em asymptotic} value of the Weyl scalar $\Psi_3$ and its gradient. In the Teukolsky formalism, that conveniently describes black hole perturbations, the Weyl scalar $\Psi_3\equiv 0$ (``transverse frame"), so that the leading order correction to $\Psi_4$ is at the next order: 
$\Psi_4=\Psi_4^0/r+O(r^{-3})$. Therefore, 
\begin{equation}
{\cal F}\sim \lim_{r\to\infty} r^2\,\Psi_4^2\sim \lim_{r\to\infty} r^2\,\left[\frac{\Psi_4^0}{r}+O(r^{-3})\right]^2\sim (\Psi_4^0)^2\,[1+O(r^{-2})]\, .
\end{equation}

We therefore are motivated to introduce the ansatz 
\begin{equation}
{\dot E}={\dot E}_{\infty}[1-A(m\lambdabar / r)^2] 
\label{ansatz}
\end{equation}
where $\lambdabar:=\lambda /(2\pi)=(r_0^{3/2}+aM^{1/2})/(mM^{1/2})$, and $r_0$ is the Boyer--Lindquist radius of the orbit. Notice that $m\lambdabar=\Omega^{-1}$, so that our ansatz can be written as ${\dot E}={\dot E}_{\infty}[1-A(\Omega r)^{-2}] $. Note that the expansion parameter $(\Omega r)^{-2}\ll 1$, as the field is evaluated far from the radiating system, specifically outside the ``light cylinder." 
This ansatz is also suggested by Fig.~1 of Ref.~\cite{burko-_khanna_07} and by Table VI of Ref.~\cite{pranesh}, where a best fit was done to the ansatz ${\dot E}_{\rm SKH}={\dot E}_{\infty}[1-q(r_0 / r)^p]$, and the free parameter $p$ was found numerically to be rather close to 2 (with deviations of $\lesssim 3\%$)  for a large range of orbits. 

We test our ansatz by fitting the outgoing flux of energy for a number of circular and equatorial orbits around a Kerr black hole as detected at a sequence of distances from the radiating system to the ansatz (\ref{ansatz}). In Table \ref{table1} we show the outgoing fluxes from particles in circular orbits around a Kerr black hole, taken at a sequence of distances. 
Table \ref{table2} shows the squared correlation coefficient for the various cases we have checked. In all cases the high value of the correlation coefficient corroborates our ansatz. In Table \ref{table3} we show the values of the free parameter $A$ for the same cases. Lastly, in Table \ref{table4} we compare the flux to infinity based on the ansatz (\ref{ansatz}) to the flux obtained from a frequency-domain calculation. Notably, the associated errors appear to be larger for $m=4,5$ than for $m=2,3$. We attribute these higher errors to the lower values of the flux for high values of $m$. Specifically, to obtain the flux with the time domain one needs to first subtract the ``flux" due to the spurious radiation associated with the initial time of the simulation \cite{burko-_khanna_07}. As the fluxes become smaller, this subtraction becomes less accurate.

\begin{table}[h]\label{table1}
 \caption{Fluxes per unit mass extracted from the time-domain code at a sequence of radii on the numerical grid. ${\dot E}_R$ is the flux per unit mass measured at a radius $RM$. For $|m|=2,3$ the data are identical to table V of Ref.~\cite{pranesh}. Notice that in \cite{pranesh} (and also tables I,II of \cite{burko-_khanna_07}) each value of $|m|$ is in fact the sum of the contributions of $m$ and $-m$. }
  \centering
     \begin{tabular}{|c|c|c||c|c|c|c|c|c||} \hline
   $|m|$ & $r_0/M$ & $a/M$ & ${\dot E}_{100}$ & ${\dot E}_{200}$ & ${\dot E}_{300}$ & ${\dot E}_{400}$ & ${\dot E}_{500}$ & ${\dot E}_{600}$   \cr \hline \hline
   1 & 4.0 & 0.99 & $1.2779\times 10^{-6}$ & $1.3049\times 10^{-6}$ & $1.3107\times 10^{-6}$  & $1.3130\times 10^{-6}$ & $1.3143\times 10^{-6}$ &  $1.3150\times 10^{-6}$  \cr \hline
   2 & 4.0 & 0.99 & $1.2284\times 10^{-3}$ & $1.2341\times 10^{-3}$ & $1.2351\times 10^{-3}$  & $1.2355\times 10^{-3}$ & $1.2356\times 10^{-3}$ & $1.2357\times 10^{-3}$  \cr \hline
   3 & 4.0 & 0.99 &$2.9481\times 10^{-4}$ & $2.9639\times 10^{-4}$ &  $2.9667\times 10^{-4}$ & $2.9677\times 10^{-4}$ & $2.9681\times 10^{-4}$ & $2.9682\times 10^{-4}$  \cr \hline
   4 & 4.0 & 0.99 & $8.3615\times 10^{-5}$ & $8.4525\times 10^{-5}$ &  $8.4769\times 10^{-5}$ & $8.4865\times 10^{-5}$ & $8.4924\times 10^{-5}$ & $8.4955\times 10^{-5}$  \cr \hline
   5 & 4.0 & 0.99 & $2.6586\times 10^{-5}$ & $2.6875\times 10^{-5}$ &  $2.6949\times 10^{-5}$ & $2.6982\times 10^{-5}$ & $2.7000\times 10^{-5}$ &  $2.7015\times 10^{-5}$  \cr \hline
    \hline

     1 & 10 & 0.90 & $1.9565\times 10^{-8}$ & $2.4481\times 10^{-8}$ & $2.5424\times 10^{-8}$  & $2.5735\times 10^{-8}$ & $2.5888\times 10^{-8}$ &  $2.5968\times 10^{-8}$  \cr \hline
   2 & 10 & 0.90 & $2.0865\times 10^{-5}$ & $2.1965\times 10^{-5}$ & $2.2161\times 10^{-5}$  & $2.2228\times 10^{-5}$ & $2.2259\times 10^{-5}$ &  $2.2275\times 10^{-5}$  \cr \hline
   3 & 10 & 0.90 &$2.3396\times 10^{-6}$ & $2.4794\times 10^{-6}$ &  $2.5043\times 10^{-6}$ & $2.5128\times 10^{-6}$ & $2.5167\times 10^{-6}$ & $2.5188\times 10^{-6}$  \cr \hline
   4 & 10 & 0.90 & $3.2046\times 10^{-7}$ & $3.4170\times 10^{-7}$ &  $3.4576\times 10^{-7}$ & $3.4723\times 10^{-7}$ & $3.4796\times 10^{-7}$ & $3.4836\times 10^{-7}$  \cr \hline
   5 & 10 & 0.90 & $4.8313\times 10^{-8}$ & $5.1497\times 10^{-8}$ &  $5.2112\times 10^{-8}$ & $5.2330\times 10^{-8}$ & $5.2438\times 10^{-8}$ &  $5.2500\times 10^{-8}$  \cr \hline
    \hline

    1 & 10 & 0.99 & $1.6454\times 10^{-8}$ & $2.0725\times 10^{-8}$ & $2.1513\times 10^{-8}$  & $2.1778\times 10^{-8}$ & $2.1907\times 10^{-8}$ &  $2.1976\times 10^{-8}$  \cr \hline
   2 & 10 & 0.99 & $2.0516\times 10^{-5}$ & $2.1605\times 10^{-5}$ & $2.1799\times 10^{-5}$  & $2.1884\times 10^{-5}$ & $2.1914\times 10^{-5}$ & $2.1931\times 10^{-5}$  \cr \hline
   3 & 10 & 0.99 &$2.2889\times 10^{-6}$ & $2.4279\times 10^{-6}$ &  $2.4526\times 10^{-6}$ & $2.4610\times 10^{-6}$ & $2.4629\times 10^{-6}$ & $2.4670\times 10^{-6}$  \cr \hline
   4 & 10 & 0.99 & $3.1481\times 10^{-7}$ & $3.3596\times 10^{-7}$ &  $3.3998\times 10^{-7}$ & $3.4144\times 10^{-7}$ & $3.4208\times 10^{-7}$ & $3.4329\times 10^{-7}$  \cr \hline
   5 & 10 & 0.99 & $4.7238\times 10^{-8}$ & $5.0395\times 10^{-8}$ &  $5.1012\times 10^{-8}$ & $5.1174\times 10^{-8}$ & $5.1122\times 10^{-8}$ &  $5.1182\times 10^{-8}$  \cr \hline
    \hline
 
    1 & 12 & 0.0 & $1.9126\times 10^{-8}$ & $2.8001\times 10^{-8}$ & $2.9932\times 10^{-8}$  & $3.0499\times 10^{-8}$ & $3.0772\times 10^{-8}$ &  $3.0928\times 10^{-8}$  \cr \hline
   2 & 12 & 0.0 & $0.9792\times 10^{-5}$ & $1.0628\times 10^{-5}$ & $1.0777\times 10^{-5}$  & $1.0827\times 10^{-5}$ & $1.0850\times 10^{-5}$ & $1.0862\times 10^{-5}$  \cr \hline
   3 & 12 & 0.0 &$0.9769\times 10^{-6}$ & $1.0663\times 10^{-6}$ &  $1.0825\times 10^{-6}$ & $1.0881\times 10^{-6}$ & $1.0906\times 10^{-6}$ & $1.0920\times 10^{-6}$  \cr \hline
   4 & 12 & 0.0 & $1.1883\times 10^{-7}$ & $1.3096\times 10^{-7}$ &  $1.3321\times 10^{-7}$ & $1.3401\times 10^{-7}$ & $1.3440\times 10^{-7}$ & $1.3462\times 10^{-7}$  \cr \hline
   5 & 12 & 0.0 & $1.5953\times 10^{-8}$ & $1.7590\times 10^{-8}$ &  $1.7874\times 10^{-8}$ & $1.7982\times 10^{-8}$ & $1.8033\times 10^{-8}$ &  $1.8062\times 10^{-8}$  \cr \hline
    \hline

   \end{tabular}
\label{table1}
\end{table}

\begin{table}[h]
 \caption{Correlation coefficient $R^2$ for the Ansatz. The roman numerals list the four orbits considered. I: $r_0=4M$, $a=0.99M$. II: $r_0=10M$, $a=0.9M$. III: $r_0=10M$, $a=0.99M$. IV: $r_0=12M$, $a=0$. Notice we have changed the order of the orbits compared with previous paper, to arrange them in ascending order of wavelengths. For all cases, the fit is done with 6 data points, at values of $r/M=100,200,300,400,500,600$. (The case II,4 has a different extraction locations.)}
  \centering
     \begin{tabular}{|c||c|c|c|c|c|} \hline
   {\bf m mode} & 1 & 2 & 3 & 4 & 5 
   \cr \hline \hline
   I & 0.998976026 & 0.99998427 & 0.999899201 & 0.9956862 & 0.99507507  \cr \hline
   II & 0.99999139 & 0.99996189 & 0.99996031 & 0.999946517 & 0.99994224  \cr \hline
   III & 0.99999644 & 0.99996171 & 0.99995783 & 0.998752311 &  0.99997298 \cr \hline
    IV & 0.99978769 & 0.99994700 & 0.99997873 &  0.99999386 & 0.99997505   \cr \hline \hline
   \end{tabular}
\label{table2}
\end{table}

\begin{table}[h]
 \caption{Best-fit value for the parameter $A$. The roman numerals designate the same orbits as in Table \ref{table1}. The bracketed number denotes the uncertainly in the last figure.}
  \centering
     \begin{tabular}{|c||c|c|c|c|c|} \hline
   {\bf m mode} & 1 & 2 & 3 & 4 & 5 
   \cr \hline \hline
   I & 3.5458(2) & 0.75753(2) & 0.86647(5) & 1.9655(3) & 1.9622(4)  \cr \hline
   II & 2.3812(2) & 0.6156(4) & 0.6915(4) & 0.7751(5) & 0.7716(5)  \cr \hline
   III & 2.4132(1) & 0.6224(4) & 0.6976(4) & 0.799(2) &  0.7734(5) \cr \hline
   IV & 2.250(2) & 0.5857(7) & 0.6287(5) & 0.6950(3) &  0.6917(5) \cr \hline \hline
   \end{tabular}
\label{table3}
\end{table}

\begin{table}[h]\label{table4}
 \caption{Comparison of frequency-domain fluxes with time-domain fluxes extrapolated to infinty based on the ansatz (\ref{ansatz}). }
  \centering
     \begin{tabular}{|c|c|c||c|c|c|} \hline
   $|m|$ & $r_0/M$ & $a/M$ & ${\dot E}_{\infty}$ & ${\dot E}_{\rm FD}$ & $({\dot E}_{\infty}-{\dot E}_{FD})/{\dot E}_{FD}$  \cr \hline \hline
   1 & 4.0 & 0.99 &  $1.3153\times 10^{-6}$ & $1.3403\times 10^{-6}$ & -0.0187 \cr \hline
   2 & 4.0 & 0.99 & $1.2359\times 10^{-3}$ & $1.2418\times 10^{-3}$ & -0.0047 \cr \hline
   3 & 4.0 & 0.99 & $2.9689\times 10^{-4}$  & $2.9621\times 10^{-4}$ & 0.0023 \cr \hline
   4 & 4.0 & 0.99 & $8.4995\times 10^{-5}$  & $8.6330\times 10^{-5}$ & -0.0155 \cr \hline
   5 & 4.0 & 0.99 & $2.7024\times 10^{-5}$ &   $2.7162\times 10^{-5}$ & -0.0051 \cr \hline
    \hline

     1 & 10 & 0.90 & $2.6147\times 10^{-8}$  &  $2.6555\times 10^{-8}$ & -0.0154 \cr \hline
   2 & 10 & 0.90 & $2.2320\times 10^{-5}$  &  $2.2281\times 10^{-5}$ & 0.0018 \cr \hline
   3 & 10 & 0.90 &$2.5245\times 10^{-6}$  & $2.5221\times 10^{-6}$ & 0.0010 \cr \hline
   4 & 10 & 0.90 & $3.4902\times 10^{-7}$ & $3.5345\times 10^{-7}$ & -0.0125 \cr \hline
   5 & 10 & 0.90 & $5.2599\times 10^{-8}$  &  $5.3255\times 10^{-8}$ & -0.0123 \cr \hline
    \hline

    1 & 10 & 0.99 & $2.2138\times 10^{-8}$  &  $2.2503\times 10^{-8}$ & -0.0162 \cr \hline
   2 & 10 & 0.99 & $2.1969\times 10^{-5}$ & $2.1974\times 10^{-5}$ & -0.0002 \cr \hline
   3 & 10 & 0.99 &$2.4726\times 10^{-6}$ & $2.4709\times 10^{-6}$  & 0.0007 \cr \hline
   4 & 10 & 0.99 & $3.4388\times 10^{-7}$  & $3.4467\times 10^{-7}$ & -0.0023 \cr \hline
   5 & 10 & 0.99 & $5.1469\times 10^{-8}$  &  $5.1687\times 10^{-8}$  & -0.0042 \cr \hline
    \hline
 
    1 & 12 & 0.0 & $3.1227\times 10^{-8}$  &  $3.1456\times 10^{-8}$ & -0.0073 \cr \hline
   2 & 12 & 0.0 & $1.0897\times 10^{-5}$ &  $1.0861\times 10^{-5}$ & 0.0033 \cr \hline
   3 & 12 & 0.0 & $1.0956\times 10^{-6}$ &  $1.0945\times 10^{-6}$ & 0.0010 \cr \hline
   4 & 12 & 0.0 & $1.3503\times 10^{-7}$ & $1.3658\times 10^{-7}$ & -0.0114 \cr \hline
   5 & 12 & 0.0 & $1.8121\times 10^{-8}$ &  $1.8317\times 10^{-8}$  & -0.0107 \cr \hline
    \hline

   \end{tabular}
\label{table4}
\end{table}

\end{appendix}


\begin{thebibliography}{99}

\bibitem{gair08} J.~R.~Gair, C.~Li, and I.~Mandel, Phys.~Rev.~D{\bf 77}, 024035 (2008) and references cited therein.
\bibitem{gair04} J.~R.~Gair, L.~Barack, T.~Creighton, C.~Cutler, S.~L.~Larson, E.~S.~Phinney, and M.~Vallisneri, Class.~Quantum Grav.~{\bf 21}, S1595 (2004).
\bibitem{glampedakis02} K.~Glampedakis and D.~Kennefick, Phys.~Rev.~D {\bf 66}, 044002 (2002).
\bibitem{hughes00} S.~A.~Hughes, Phys.~Rev.~D {\bf 61}, 084004 (2000).
\bibitem{lopez_aleman} R.~Lopez--Aleman, G.~Khanna, and J.~Pullin, Class.~Quantum Grav.~{\bf 20}, 3259 (2003).
\bibitem{khanna04} G.~Khanna, Phys.~Rev.~D {\bf 69}, 024016 (2004).
\bibitem{burko-_khanna_07} L.~M.~Burko and G.~Khanna, Europhysics Letters {\bf 78}, 60005 (2007) [arXiv:gr-qc/0609002].
\bibitem{pranesh} P.~A.~Sundararajan, G.~Khanna, and S.~A.~Hughes, Phys.~Rev.~D (in press) [arXiv:gr-qc/0703028].
\bibitem{pranesh2} P.~A.~Sundararajan, G.~Khanna, S.~A.~Hughes, and S.~Drasco, arXiv:0803.0317. 
\bibitem{schmidt} W.~Schmidt, Class.~Quantum Grav.~{\bf 19}, 2743 (2002).
\bibitem{drasco-hughes} S.~Drasco and S.A.~Hughes, Phys.~Rev.~D {\bf 69}, 044015 (2004).
\bibitem{finn_thorne_00} L.~S.~Finn and K.~S.~Thorne, Phys.~Rev.~D {\bf 62}, 124021 (2000).
\bibitem{peters_mathews_63} P.~C.~Peters and J.~Mathews, Phys.~Rev.~{\bf 131}, 435 (1963).
\bibitem{finn} L.~S.~Finn, private communication.
\bibitem{hughes-2000} S.A.~Hughes, Phys.~Rev.~D {\bf 61}, 084004 (2000); Erratum-ibid. D {\bf 63} 049902 (2001).
\bibitem{num1} C.~Cutler, L.~S.~Finn, E.~Poisson, and G.~J.~Sussman, Phys.~Rev.~D {\bf 47}, 1511 (1993).
\bibitem{s-n} M.~Sasaki and T.~Nakamura, Prog.~Theor.~Phys.~{\bf 67}, 1788 (1982).
\bibitem{num_rec} W.H.~Press, S.A.~Teukolsky, W.T.~Vetterling, and B.P.~Flannery, {\em Numerical Recipes: The Art of Scientific Computing}, Third Ed.~(Cambridge University Press, Cambridge, 2007).
\bibitem{stephani} H.~Stephani, {\em Relativity: An Introduction to Special and General Relativity}, 3$^{\rm rd}$ ed., (Cambridge University Press, Cambridge, 2004). 
\bibitem{stewart} J.~Stewart, {\em Advanced General Relativity} (Cambridge University Press, Cambridge, 1990).
\bibitem{newman_unti_62} E.~T.~Newman and T.~Unti, J.~Math.~Phys.~{\bf 3}, 891 (1962).

\end{thebibliography}
\end{document}